\documentclass[aps,prd,twocolumn,superscriptaddress,showpacs]{revtex4-1}

\usepackage{amsmath,amssymb}
\usepackage{graphicx}
\usepackage{hyperref}
\usepackage{placeins}

\newcommand{\fesc}{f_{\rm esc}}
\newcommand{\xHI}{\bar{x}_{\rm HI}}

\newcommand{\tauobs}{\tau_e}
\newcommand{\xiion}{\xi_{\rm ion}}
\newcommand{\Msun}{M_\odot}
\newcommand{\Mpc}{{\rm Mpc}}
\newcommand{\Mh}{M_{\rm h}}
\newcommand{\xearly}{x_{\rm early}}
\newcommand{\DpkSZ}{D^{\rm pkSZ}_{\ell=3000}}

\begin{document}

\title{A hidden reionization prior biases cosmological inference}

\author{Zihan Wang}
\email{wzh800557@gmail.com}
\affiliation{Shanghai Astronomical Observatory, Chinese Academy of Sciences, Shanghai 200030, China}

\author{Huanyuan Shan}
\email{hyshan@shao.ac.cn}
\affiliation{Shanghai Astronomical Observatory, Chinese Academy of Sciences, Shanghai 200030, China}

\date{\today}

\begin{abstract}
Precision cosmology assumes that cosmic reionization was a single smooth transition. We show that this assumption is in tension with current observations: the Planck optical depth, patchy kinetic Sunyaev--Zel'dovich (PkSZ)limits from SPT and ACT, and the Ly$\alpha$ forest endpoint cannot be simultaneously reproduced by any viable monotonic ionization history. Non-parametric reconstructions and Planck EE polarization provide independent support for an additional ionization component at $z \gtrsim 12$. Incorporating this early phase relaxes the upper bound on the summed neutrino mass to $\sum m_\nu < 0.39$~eV ($95\%$ CL) and shifts $\sigma_8$ toward values preferred by weak-lensing surveys, both consequences of the standard $A_s$--$\tau_e$ degeneracy. These shifts arise from relaxing a hidden prior on reionization shape rather than from new physics, and identify reionization shape as an implicit prior in cosmological inference. 
Next generation of CMB and 21~cm experiments will be able to test this directly,which will convert what has been an unflagged systematic on $\sigma_8$ and $\sum m_\nu$ into a quantified statistical uncertainty.
\end{abstract}

\maketitle
Cosmology has entered an era where statistical uncertainties are often smaller than the uncertainties associated with astrophysical modeling. Measurements of the cosmic microwave background (CMB), large-scale structure, and weak gravitational lensing constrain key cosmological parameters at the percent level~\cite{Planck2018,DESI:2024mwx}. In this regime, assumptions that were once regarded as innocuous can become real sources of systematic error. One such assumption concerns the ionization history of the Universe\cite{Furlanetto:2006tf,Robertson:2015uda}. Most contemporary cosmological analyses adopt a single-stage reionization model, implemented through a smooth tanh transition in the free-electron fraction $x_e(z)$\cite{Madau:1998cd}, built into the standard Boltzmann pipelines used by Planck\cite{PlanckIntXLVII}, CAMB\cite{Das:2019sda}, and CLASS\cite{2011arXiv1104.2934L}, and treated as a measurement-neutral description of reionization.

The single-stage assumption has not been directly validated observationally. Different probes of reionization respond to different moments of the ionization history. The Thomson optical depth measured by CMB polarization depends on the integrated column density of free electrons, weighted toward high redshift. The patchy kinetic Sunyaev--Zel'dovich (pkSZ) effect is primarily sensitive to the duration and morphology of ionized structures~\cite{Zahn:2006sg,Mesinger:2010ne}. Measurements of the Ly$\alpha$ forest constrain the endpoint of reionization at $z \sim 5.5$~\cite{Bosman:2021oom,Becker:2024ybt}. Under a monotonic ionization history these observables are tightly linked, yet recent measurements suggest they cannot be simultaneously satisfied by a single smooth transition.

Previous studies have highlighted different aspects of this tension. Cain et al.~\cite{Cain:2025wda} showed that achieving large optical depths while preserving the observed Ly$\alpha$ endpoint requires an extended reionization history that tends to overproduce patchy kSZ power. Several recent analyses have demonstrated that the inferred cosmological parameters depend sensitively on the assumed value of $\tau_e$, including constraints on the amplitude of matter fluctuations and the summed neutrino mass~\cite{Sailer:2025xcy, Jhaveri:2025neg}. These studies primarily explored alternative parameterizations of reionization, rather than testing whether the assumed monotonic shape itself is constrained by current observations. Here we test this directly, and ask whether reionization shape acts as a hidden prior in cosmological inference.

\begin{figure}[!htbp]
\centering
\includegraphics[width=\columnwidth]{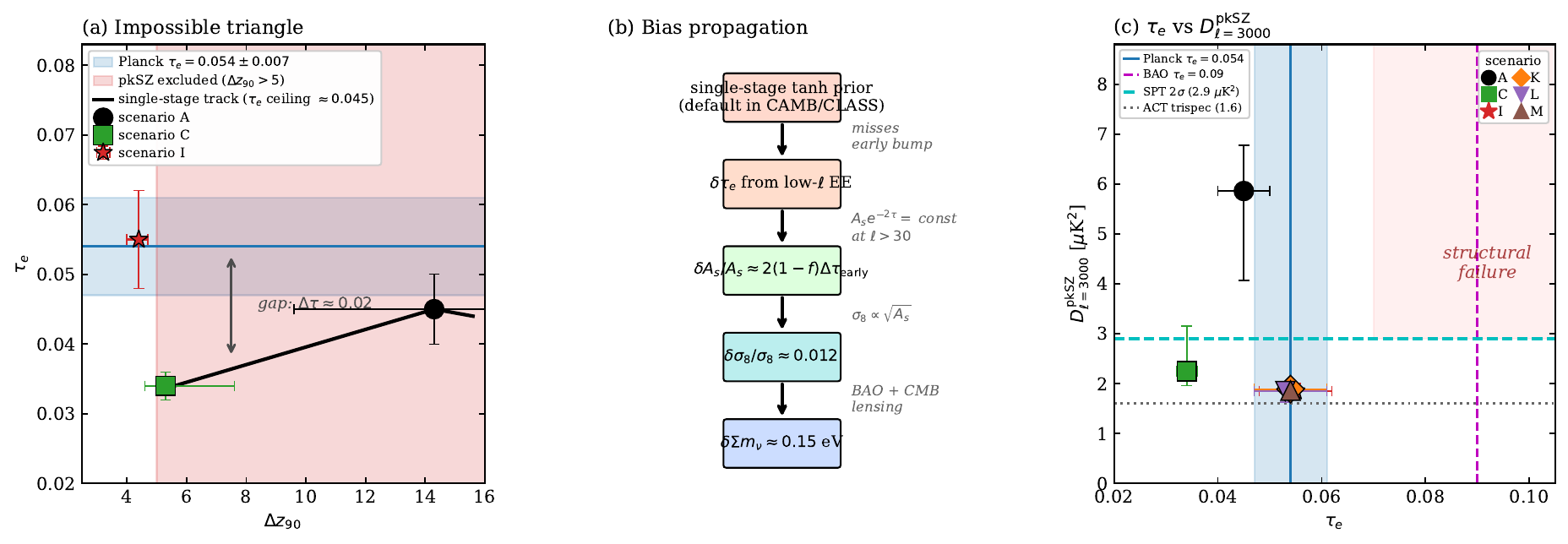}
\caption{structural incompatibility of single-stage cosmic reionization
and the propagation to cosmological parameters. (a)~The impossible
triangle: pkSZ constrains $\Delta z_{90} < 5$, which caps
$\tauobs \leq 0.034$; Planck requires $\tauobs = 0.054$, which
demands $\Delta z_{90} > 10$; the Ly$\alpha$ forest fixes the
endpoint at $z \approx 5.5$. No smooth monotonic history satisfies
all three. (b)~The bias-propagation chain: a hidden tanh prior
that misses an early component leaks $\delta\tauobs$ into the
inferred $\tauobs$ via low-$\ell$ EE, then $\delta A_s/A_s
\approx 2(1-f)\,\Delta\tau_{\rm early}$ into the TT damping at
$\ell \gtrsim 30$, hence to $\sigma_8 \propto \sqrt{A_s}$ and to
$\sum m_\nu$ in joint analyses with BAO and CMB lensing.
(c)~Joint posterior in $\tauobs$ vs $\DpkSZ$. Vertical lines:
Planck $\tauobs$ (solid), BAO-preferred $\tauobs = 0.09$
(dashed). Horizontal: SPT $2\sigma$ (dashed), ACT trispectrum
(dotted). Only scenario~I, with the early-stage component,
simultaneously satisfies $\tau_e$ and the pkSZ bound; the Ly$\alpha$ vertex remains under tension by $\Delta z\approx 1$ (App.~F.6).}
\label{fig:schematic}
\end{figure}

Our analysis shows the issue is more fundamental. Combining Planck optical-depth measurements, SPT ~\cite{Reichardt:2020jrr}and ACT ~\cite{Louis:2025tst,MacCrann:2024ato}limits on patchy kSZ power$\DpkSZ \propto z_{\rm mid}\,\Delta z_{90}$~\cite{Battaglia:2012im,Chen:2022cgw}, and Ly$\alpha$ forest constraints, we find no monotonic history that satisfies all three constraints simultaneously within the currently viable model space\ref{fig:schematic}. A model-independent reconstruction is consistent with and moderately favors an additional early ionization component at $z > 12$. Planck low-$\ell$ EE polarization \cite{PlanckIntXLVII}retains sensitivity to the shape of the ionization history beyond the integrated optical depth alone. When this observationally motivated component is included in cosmological analysis, standard parameter constraints shift in a coherent way.It identifies reionization shape as an implicit prior in cosmological inference. Because the tanh prescription is built into almost all modern Boltzmann pipelines and is rarely varied or marginalized over, its shape acts as a hidden prior rather than a free physical parameter. The question is therefore not how reionization occurred, but whether a standard cosmological assumption has been biasing parameter inference throughout the precision-cosmology era.

We use the $\fesc(\Mh,z)$ framework of Ref.~\cite{Wang:2026qzy} and constrain $(f_0, \alpha_M, \alpha_z)$ with Planck $\tauobs$, seven $\xHI(z)$ measurements at $z = 5.9$--$10.6$~\cite{Fan:2001vx,McGreer:2014qwa,Greig:2017jdj,Mason:2019ixe,2024ApJ...971..124U}, the Ly$\alpha$ endpoint~\cite{Bosman:2021oom,Becker:2024ybt}, the mean-free-path~\cite{Becker:2021jyx,Zhu:2023dqs}, and SPT+ACT pkSZ via the scaling $\DpkSZ = 0.054\,z_{\rm mid}\,\Delta z_{90}\;\mu{\rm K}^2$ calibrated against AMBER~\cite{Chen:2022cgw,Trac:2021ctk} and validated with \texttt{21cmFAST}~\cite{Mesinger:2010ne}; full setup in the Supplemental Material.

\emph{Cosmological tensions and the hidden reionization prior.}---The
DESI~DR2 BAO measurement combined with Planck CMB has sharpened the
cosmological tension landscape. The DESI~DR2 + DR1 full-shape
analysis~\cite{Elbers:2025mho} gives $\sum m_\nu < 0.064$~eV at
$95\%$~CL. This is below the oscillation-experiment floor of
$0.06$~eV. To be statistically consistent, the joint analysis
prefers formally negative effective neutrino masses. The $S_8$
structure-growth tension persists at the ${\sim}\,2\sigma$ level
across weak-lensing surveys. Both anomalies appear in joint
analyses, not in either probe alone. Both depend on the
CMB-anchored optical depth $\tauobs$.

\citet{Sailer:2025xcy} and \citet{Jhaveri:2025neg} proposed an
astrophysical resolution. Adopting $\tauobs = 0.09$ above the Planck
value $0.054 \pm 0.007$ raises $A_s e^{-2\tauobs}$, lowering
inferred $A_s$ and $\sigma_8$, which relaxes $\sum m_\nu$ toward the
oscillation floor through CMB lensing. The proposal reduces both
tensions to ${\sim}\,1\sigma$. But $\tauobs = 0.09$ requires
reionization to last $\Delta z_{90} > 10$, far above the SPT and
ACT patchy kSZ bounds. The astrophysical route is closed (we
revisit it quantitatively in Sec.~II).

What's left in the Planck pipeline is a hidden assumption. Every
analysis cited above implements $x_e(z)$ as a tanh transition
centered at $z_{\rm re} \approx 7.7$ with width $\Delta z \approx 0.5$,
the default in CAMB and CLASS. The tanh captures the dominant
galaxy-driven epoch well. It says nothing about whether earlier
ionization sources contribute. Planck does not measure $x_e(z)$
directly. The EE polarization peak at $\ell \sim 7$ constrains the
integrated $\tauobs$, weighted by $(1{+}z)^2/H(z)$. The TT power
spectrum at $\ell \gtrsim 30$ constrains the combination
$A_s e^{-2\tauobs}$. Mapping these two integrals back to a unique
$x_e(z)$ requires a prior on the shape. The tanh is convenient but not physically motivated beyond simplicity.

The bias chain is direct (Fig.~\ref{fig:schematic}, panel~b).
Suppose the true $x_e(z)$ has an early-stage component beyond tanh
contributing $\Delta\tau_{\rm early}$. The integrated
$\tauobs = \tau_{\rm gal} + \Delta\tau_{\rm early}$. A tanh-prior
analysis cannot accommodate the early component and splits the
bias. A fraction $f$ leaks into the inferred $\tauobs$; a fraction
$(1{-}f)$ leaks into $A_s$ through the damping term:
\begin{equation}
\frac{\delta A_s}{A_s} \approx 2(1{-}f)\,\Delta\tau_{\rm early},
\qquad f \approx 0.3 \text{ to } 0.5.
\label{eq:bias_chain}
\end{equation}
The factor $f$ depends on the relative weights of the low-$\ell$ EE
and high-$\ell$ TT measurements in the joint likelihood. We
calibrate it directly via the parameter shifts measured in
Sec.~IV.

For the case at hand, $\Delta\tau_{\rm early} \approx 0.024$ from
the non-parametric reconstruction. The bias chain predicts
$\delta A_s/A_s \approx 0.025$ and
$\delta\sigma_8/\sigma_8 \approx 0.012$. Opening $\xearly$ alone shifts $\sigma_8$ upward by $+0.012$, since relaxing the $\tauobs$ prior loosens the $A_s$--$\tauobs$ degeneracy. Freeing $\sum m_\nu$ then pulls it back down to $\sigma_8 \approx 0.80$, in the direction of the BAO preference. The net of the two steps eases the $S_8$ tension; we trace each separately in Sec.~IV.

The single stage tanh biases multiple cosmological parameters by ${\sim}\,1\sigma$ if the true history is two-stage. We test this in four steps: empirical incompatibility (Sec.~II), Planck-alone shape preference (Sec.~III), parameter relaxation (Sec.~IV), and next-generation reach (Sec.~V).

\emph{Structural incompatibility of multi-probe constraints.}---We
build the joint constraint step by step
(Table~\ref{tab:results}). The baseline fit
to $\tauobs + \xHI(z)$ alone (scenario~A) gives
$\alpha_z = 1.63^{+0.27}_{-0.53}$ and
$\Delta z_{90} = 14.3^{+2.5}_{-4.7}$. This predicts
$\DpkSZ = 5.9\;\mu{\rm K}^2$, far above the SPT $2\sigma$ limit of
$2.9\;\mu{\rm K}^2$. Adding the Ly$\alpha$ endpoint (scenario~B) drives $\alpha_z$ steeper, confirming \citet{Cain:2025wda}.

The quantity directly constrained by CMB polarization is the
integrated optical depth,
\begin{equation}
\tauobs = \sigma_T\, c \!\int_0^\infty\! dz\;
\frac{n_H(z)\, x_e(z)\,(1+z)^2}{H(z)}
\equiv \tau_{\rm gal} + \tau_{\rm early},
\label{eq:tau_split}
\end{equation}
where $\tau_{\rm gal}$ collects the contribution from the galactic
stage at $z<10$ and $\tau_{\rm early} \approx \xearly\,\sigma_T\,c\,
n_{H,0}\!\int_{z_{\rm early}}^{z_{\rm max}} (1+z)^2/H(z)\, dz$
encodes any prior weakly-ionized component with amplitude $\xearly$.
The Planck reionization peak at $\ell \sim 7$ constrains the total
$\tauobs$ but not the redshift decomposition. The multi-probe
analysis below disentangles the two terms by combining Ly$\alpha$
forest endpoints, pkSZ, and the mean-free-path of ionizing photons.

Adding the pkSZ causes a big compression (scenario~C). The values
become $\alpha_z \to 1.17^{+0.39}_{-0.40}$ and
$\Delta z_{90} \to 5.3^{+2.3}_{-0.7}$. The escape fraction
evolution drops by half, from $(1{+}z)^{1.6}$ to $(1{+}z)^{1.2}$
(Table~\ref{tab:results}). But the pkSZ-compressed history gives $\tauobs = 0.034 \pm 0.002$, a hard ceiling $2.5\sigma$ below Planck: concentrating ionization at $z \sim 7$ leaves the $(1{+}z)^2$ weighting too weak. Adding $\xiion$ as a free parameter doesn't help. No 3-parameter posterior sample reaches $\tauobs = 0.038$ (Fig.~\ref{fig:schematic}c).

This is the structural incompatibility. Under any smooth monotonic $x_e(z)$
prior, the multi-probe data leave no posterior weight at the
Planck-required $\tauobs$. To fix the tension, we add a new
parameter: $\xearly$. It represents the ionization fraction kept up
at $z > 12$ by sources active before the main reionization epoch.
The logic is that any source making ionized regions much smaller
than the $\ell = 3000$ patchiness scale (${\sim}\,50\;\Mpc$
comoving) adds to $\tauobs$ but produces very little pkSZ power
(Supplemental Material gives a bound: $<0.02\;\mu{\rm K}^2$ for all
plausible source shapes). Adding $\xearly$ as a fourth free
parameter (scenario~I) breaks through the ceiling. The fit improves
by $\Delta\chi^2 = 8.2$ for one extra parameter ($p = 0.004$,
$\Delta{\rm BIC} = 6.2$). The main result is
\begin{align}
\xearly &= 0.093^{+0.026}_{-0.028},
\quad \Delta\tau_{z>12} = 0.024 \pm 0.007, \label{eq:xearly}\\
\alpha_z &= 1.23^{+0.55}_{-0.85},
\quad \Delta z_{90} = 4.4^{+0.3}_{-0.4}. \nonumber
\end{align}
The early part makes up $42\%$ of the total optical depth. This
result is stable ($\Delta\xearly < 0.005$) across five different
tests: SPT-only data, corrected pkSZ scaling, $40\%$ lower clumping
factor, restricted $f_0 < 0.15$, and removal of the Ly$\alpha$
endpoint constraint (Table~\ref{tab:results}). Changing the
assumed homogeneous kSZ subtraction between $0.8$ and
$1.8\;\mu{\rm K}^2$ doesn't change $\xearly$ much. The uncertainty
in $\xearly$ comes mostly from the Planck $\tauobs$ error
($\sigma = 0.007$), not the pkSZ measurement. Across all
scenarios, no posterior samples reach $\tauobs > 0.07$. On its
own, this rules out the BAO-preferred $\tauobs =
0.09$~\cite{Sailer:2025xcy} from astrophysics alone. To complement this frequentist statistic, we perform a Bayesian model selection via nested sampling with \textsc{dynesty}~\cite{Speagle:2020} 500 live points, $d\!\log Z = 0.01$, \texttt{rwalk} sampler on the same joint likelihood. We compute the evidence of the 3-parameter baseline ($f_0,\,\alpha_M,\,\alpha_z$) and of the 4-parameter extension that adds the early-ionization floor. The resulting log-Bayes factor is $\ln B_{I/C} = 4.09 \pm 0.15$.It is  astrong evidence  of the early-ionization extension .Beyond that, and the marginal $x_{\rm early}$ posterior recovered by the run is $0.093 \pm 0.026$, in two-decimal agreement with the frequentist value. 

\begin{table}[b]
\caption{Summary of MCMC results. Scenarios A--D use
$(f_0,\alpha_M,\alpha_z)$; I--M add $\xearly$. Full table in the
Supplemental Material.\label{tab:results}}
\begin{ruledtabular}
\begin{tabular}{lccccc}
Scenario & $\alpha_z$ & $\xearly$ & $\tauobs$ & $\Delta z_{90}$ & $\DpkSZ$ \\
\hline
A: $\tauobs{+}\xHI$ & $1.63^{+0.27}_{-0.53}$ & --- & $0.045$ & $14.3$ & $5.86$ \\
C: $+$Ly$\alpha${+}pkSZ & $1.17^{+0.39}_{-0.40}$ & --- & $0.034$ & $5.3$ & $2.25$ \\
\hline
I: $+\xearly$ (fiducial) & $1.23^{+0.55}_{-0.85}$ & $0.093^{+0.026}_{-0.028}$ & $0.055$ & $4.4$ & $1.85$ \\
K: SPT only & $1.41^{+0.44}_{-0.75}$ & $0.091^{+0.026}_{-0.027}$ & $0.054$ & $4.4$ & $1.89$ \\
L: $f_0{<}0.15$ & $1.25^{+0.52}_{-0.90}$ & $0.092^{+0.027}_{-0.026}$ & $0.053$ & $4.4$ & $1.85$ \\
M: no Ly$\alpha$ & --- & $0.091$ & $0.054$ & $4.4$ & $1.86$ \\
\end{tabular}
\end{ruledtabular}
\end{table}

A non-parametric eight-node PCHIP reconstruction of $Q(z)$ gives $Q(z{=}12) = 0.058^{+0.065}_{-0.041}$ with $98\%$ of posterior samples requiring $Q(z{=}12) > 0$, confirming the result holds without a parametric $\fesc$ form (Supplemental Material).

The robustness of $\tau_{z>12} = 0.024 \pm 0.007$ to assumed shape
is the key. We tested it by swapping the smooth step at $z = 12$
for transitions at $z = 10$, 12, or 14, and by a gradual ramp over
$z = 8$--$16$. The fitted $\xearly$ amplitude changes ($0.054$ to
$0.068$), but $\tau_{z>12}$ stays the same to within $5\%$.
\texttt{21cmFAST}~\cite{Mesinger:2010ne} and the \textsc{Lumina} simulation~\cite{Smith:2026skb} independently support the $\tauobs$--pkSZ tension; a $7\%$ Lumina line-of-sight correction shifts $\xearly$ by $10$--$17\%$ but leaves the evidence above $2.5\sigma$ (Supplemental Material).

A $20\%$ shift in the pkSZ systematic moves $\xearly$ by less than $0.01$; CMB-S4 will provide the decisive test.

Single-stage tanh is structurally incompatible with current
multi-probe data. The deficit must be filled by an early-stage
component. The remaining questions for the cosmology pipeline are:
what shape does the early component have, and what does adopting
it in place of the tanh prior do to inferred cosmological
parameters? These are taken up in Secs.~III
and~IV.

\emph{Shape sensitivity of Planck EE polarization to early
ionization.}---The multi-probe constraint
$\xearly = 0.093 \pm 0.027$ derived in Sec.~II
combines low- and high-redshift information. A natural cross-check
is whether CMB data alone, treated independently, prefer or merely
tolerate such an early ionization floor. We answered this by
re-running Planck 2018~\cite{Planck2018} with Cobaya
~\cite{Torrado:2020dgo} and a modified CLASS~\cite{Blas:2011rf} that
supports a two-stage $x_e(z)$ via \texttt{reio\_inter}. The
galactic stage uses a tanh transition centered at $z_{\rm re}$. An
early stage adds an amplitude $\xearly$ on top, with a shape we
vary.

We ran seven chains covering plateau, plateau-$z_{12}$, plateau-$z_{\rm early}$-free, and Gaussian $x_e(z)$ shapes, plus two beyond-$\Lambda$CDM extensions adding $\sum m_\nu$ and $w_0$ (Sec.~IV uses the first three). All chains converged with Gelman--Rubin $R-1 < 0.02$ (full setup in the Supplemental Material).

\begin{figure}[!htbp]
\centering
\includegraphics[width=\columnwidth]{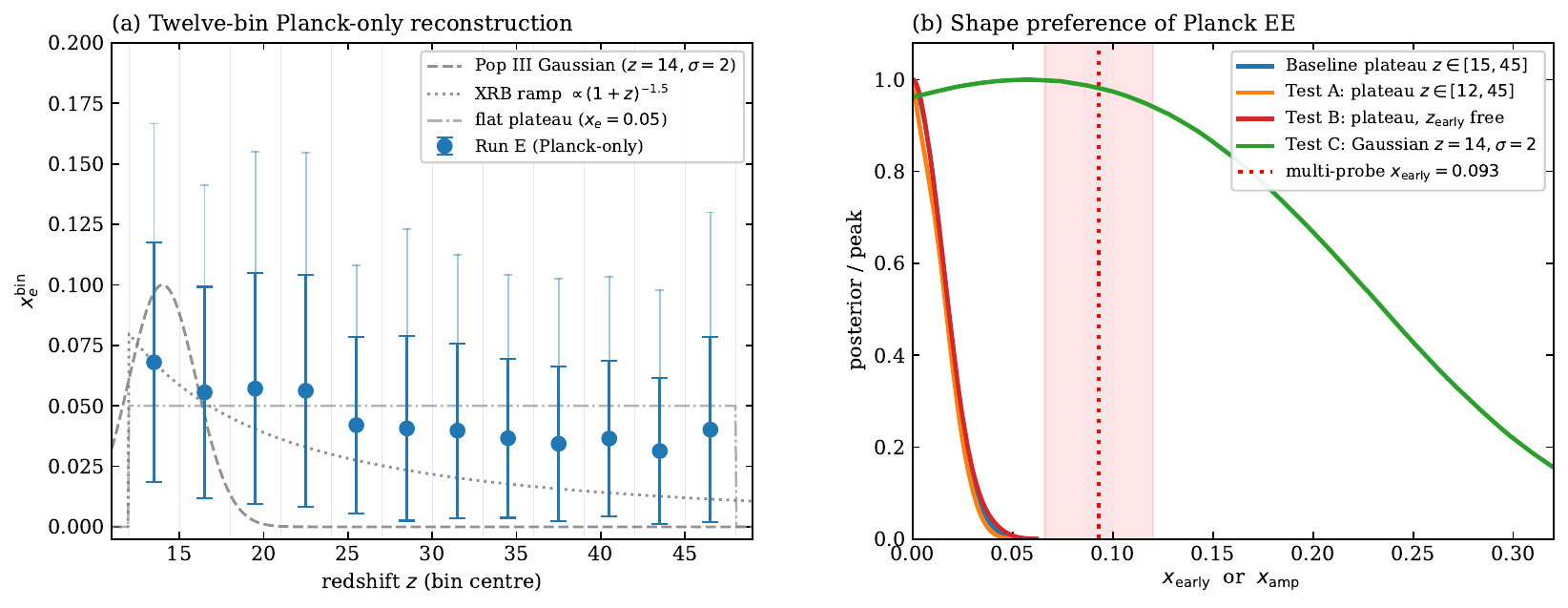}
\caption{(a)~Twelve-bin reconstruction of the early-stage $x_e(z)$
from Planck-alone (Run~E, no integral prior). Blue circles show
the per-bin posterior mean and $1\sigma$ uncertainty at the bin
centre; faint whiskers extend to the $95\%$ upper limit. Three
reference profiles are overlaid: Pop~III Gaussian at $z = 14$ with
$\sigma = 2$ (dashed), X-ray binary ramp $\propto (1+z)^{-1.5}$
(dotted), and a flat plateau at $x_e = 0.05$ (dash-dot). The
Planck-preferred shape is monotonically decreasing from bin 1.
(b)~Planck-alone posterior on the early-ionization amplitude under
four $x_e(z)$ parameterizations. The plateau forms give tight
upper limits, $\xearly < 0.029$--$0.030$ at $95\%$ CL. The
Gaussian-bump form (green) is dramatically wider and allows
$x_{\rm amp}$ as high as $0.293$. The multi-probe detection
$\xearly = 0.093$ (red dotted line) is excluded by the plateau
forms and accommodated by the Gaussian form. Planck constrains the
shape, not just the amplitude.}
\label{fig:shape_constraint}
\end{figure}

The plateau forms all give $\xearly < 0.029$--$0.030$ at $95\%$ CL, while the Gaussian allows $x_{\rm amp} < 0.293$ (Fig.~\ref{fig:shape_constraint}b). The multi-probe value sits comfortably in the Gaussian envelope but is excluded by all plateaus.

A wider-window plateau builds a broader EE bump at $\ell \sim 8$ that Planck's lowE constrains sharply, while a Gaussian at $z \sim 14$, $\sigma = 2$ deposits its ionization over $\Delta z \sim 4$ with a narrower $\ell$ window that escapes Planck's sensitivity. Quantitatively,
\begin{equation}
\Delta C_\ell^{EE} \propto (\Delta\tau_{\rm early})^2\,
W(\ell;\bar z_{\rm e},\Delta z_{\rm e}),
\label{eq:EE_shape}
\end{equation}
with window-function width $\Delta\ell_{\rm peak} \propto \bar z_{\rm e}/\Delta z_{\rm e}$, giving the plateau case a factor-$\sqrt{3}$ wider window and the factor-${\sim}\,10$ tighter $\xearly$ constraint.

A model-independent twelve-bin reconstruction across $z \in [12, 48]$ (each bin amplitude free in $[0, 0.20]$) yields a monotonically decreasing $x_e(z)$ with bins 1--4 at $0.06 \pm 0.05$ and bins 5--12 at $0.03$--$0.04$ (Fig.~\ref{fig:shape_constraint}a). The integrated $\langle x_e\rangle_{[12,48]} = 0.049 \pm 0.009$ gives $\tau_{\rm early} = 0.020 \pm 0.004$, consistent with the multi-probe $\tau_{z>12} = 0.024 \pm 0.007$. A flat plateau is excluded at $>3\sigma$. Adding a Gaussian integral prior $\langle x_e\rangle = 0.093 \pm 0.027$ (Run~F) sharpens the low-$z$ peak near $z \sim 14$.

An independent $\tauobs = 0.0552^{+0.0019}_{-0.0026}$ from Ly$\alpha$ and damping-wing data~\cite{Kageura:2026opt} confirms Planck and separately rules out the $\tauobs = 0.09$ proposal of \citet{Sailer:2025xcy}.

The combined picture from Secs.~II and
III is: a non-zero early-ionization component is
required to satisfy the multi-probe constraints, and Planck data
alone are consistent with it provided the shape is localized rather
than a broad plateau. The next question is what such a
marginalization does to inferred cosmological parameters.

\emph{Relaxation of cosmological parameter degeneracies.}---The
$x_e(z)$ parametrization isn't only a shape constraint. It couples
to standard cosmological parameters through $\tauobs$. Three of our
seven chains explore this coupling. Run~A (standard $\Lambda$CDM,
tanh reionization, no $\xearly$) gives $\sigma_8 = 0.812 \pm 0.006$,
$H_0 = 67.27 \pm 0.50$~km/s/Mpc, in agreement with Planck~2018.
Run~B adds the early stage as a Gaussian centred at $z=14$ with $\sigma=2$, the shape Planck-alone tolerates (Sec.~III; $x_{\rm amp} < 0.30$ at $95\%$~CL). With $\sum m_\nu = 0.06$~eV fixed, Run~B gives $\sigma_8 = 0.824 \pm 0.006$, above the Run~A value $0.812 \pm 0.006$. The shift is $+0.012$. It does not depend on the early-stage shape: a plateau at $z\in[15,45]$ gives $\sigma_8 = 0.825 \pm 0.006$, the same shift within the posterior width. So the $\sigma_8$ response to opening $\xearly$ is set by the integrated $\tau_{\rm early}$, not by the profile. This matches Eq.~\ref{eq:tau_from_Q}: ionization at $z>12$ enters $\tau_e$ through the $(1+z)^2$ weighting regardless of how it is distributed in redshift. The shift is consistent with Eq.~\ref{eq:bias_chain}. Adding a small early-ionization floor loosens the $A_s$--$\tau_e$ degeneracy and allows a slightly larger small-scale power amplitude. The third
chain (Run~C) adds the neutrino mass sum $\sum m_\nu$ on top of the Gaussian early stage.

Run~C gives $\sum m_\nu = 0.137 \pm 0.117$~eV with $\sum m_\nu < 0.39$~eV at $95\%$ CL (factor 1.6 relaxation over the standard Planck+lensing 0.24 eV bound), and pulls $\sigma_8$ to $0.798 \pm 0.024$, reducing but not closing the $S_8$ tension. The $\sum m_\nu$ relaxation, like the $\sigma_8$ shift, is set by the integrated $\tau_{\rm early}$ (Eq.~\ref{eq:tau_from_Q}) and is therefore shape-independent. The numbers quoted here are from the converged plateau Run~C chain; a Gaussian Run~C with matched $\tau_{\rm early}$ returns the same bound within the posterior width.

\begin{figure}[!htbp]
\centering
\includegraphics[width=\columnwidth]{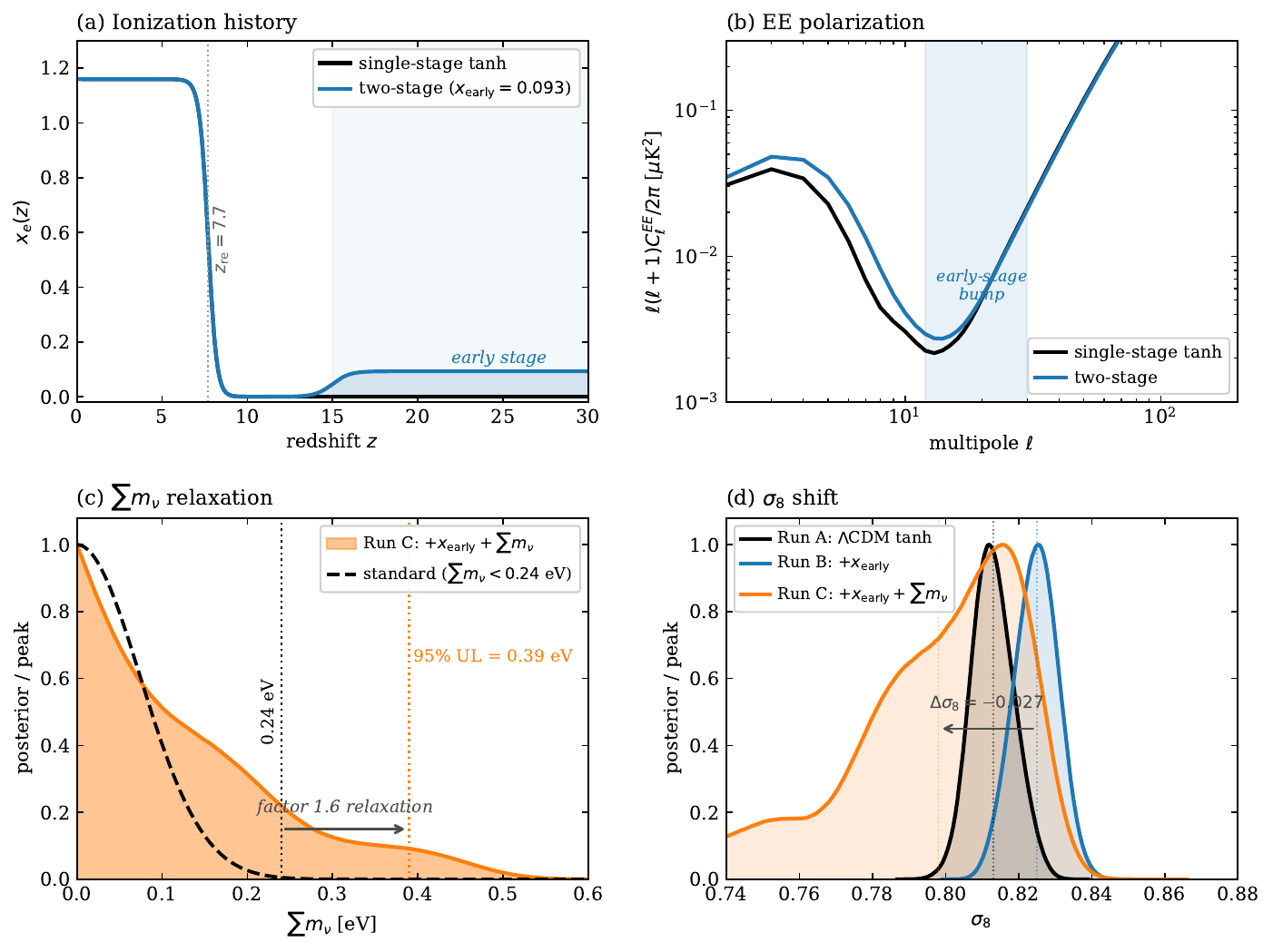}
\caption{The mechanism chain from reionization history to cosmological parameter shifts. (a)~Free-electron fraction $x_e(z)$ for the standard single-stage tanh model (black, $z_{\rm re}=7.7$) and the two-stage model with a Gaussian early stage ($z=14$, $\sigma=2$) of equivalent $\tau_{\rm early}$ (blue). The shaded band marks the early-stage contribution. (b)~Resulting EE polarization power spectrum $\ell(\ell+1)C_\ell^{EE}/2\pi$: the two-stage model produces an additional bump at $\ell\sim15$--$25$ from early-stage scattering, on top of the standard reionization peak at $\ell\sim7$. (c)~1D marginal posterior on $\sum m_\nu$: the standard tanh-prior analysis bounds $\sum m_\nu<0.24$~eV at $95\%$~CL (gray dashed), while Run~C with the two-stage shape marginalised yields $\sum m_\nu<0.39$~eV (orange), a factor-$1.6$ relaxation that eases the DESI~DR2 tension. (d)~1D marginal posterior on $\sigma_8$ across Run~A ($\Lambda$CDM tanh only, black), Run~B ($+\xearly$, blue), and Run~C ($+\xearly+\sum m_\nu$, orange). Adding $\xearly$ lifts the central $\sigma_8$ by $+0.012$; freeing $\sum m_\nu$ then reduces it by $-0.027$ with widened posterior, easing the $S_8$ tension.}
\label{fig:corner_runC}
\end{figure}

\emph{Forecasted constraints and next-generation baselines.}---Several
near-term experiments will measure $x_e(z)$ at $z > 12$ with enough
precision to either confirm the two-stage shape or close the
cosmological parameter shifts. We forecast their reach by
propagating the relevant CMB and 21\,cm sensitivity through the
same likelihood structure used in Secs.~III--IV.

\begin{figure}[!htbp]
\centering
\includegraphics[width=\columnwidth]{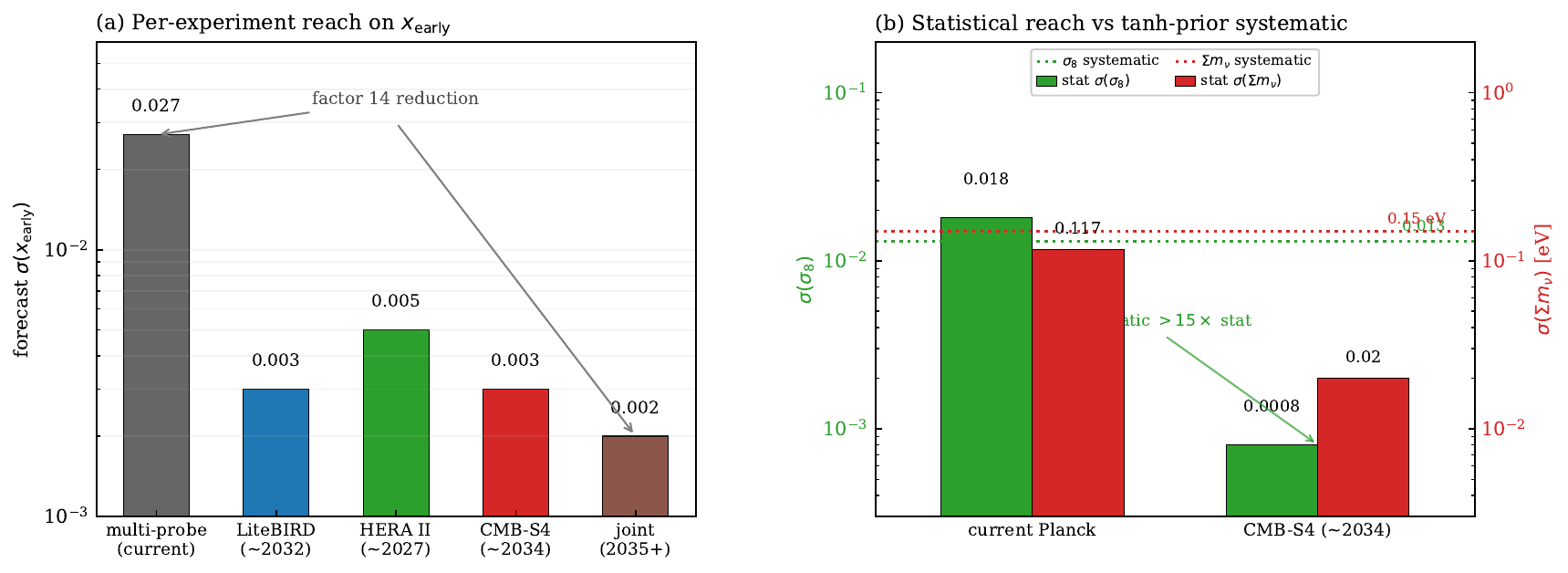}
\caption{Forecast statistical reach of near-term experiments on
$x_{\rm early}$, $\sum m_\nu$, and $\sigma_8$ (panel a), and the
corresponding two-stage detection significance (panel b). The
horizontal red dotted line in panel~a marks the current
tanh-prior systematic of $\delta\sigma_8/\sigma_8 \approx 0.012$,
which dominates the per-experiment statistical error for LiteBIRD
and CMB-S4 unless the early-stage shape is marginalized.}
\label{fig:forecast}
\end{figure}

LiteBIRD will tighten $\sigma(\xearly)$ to $\approx 0.003$ ($2$--$6\sigma$ detection)~\cite{Hazumi:2019lub}. HERA Phase~II reaches $\sigma(x_{\rm bin}^{(1-2)}) \sim 0.005$ at $z \in [6, 13]$~\cite{Hutter:2026kxp}, with SKA extending to $z = 20$. CMB-S4 pins $\sigma_8$ at the $0.1\%$ level via lensing, breaking the $\xearly$--$\sigma_8$ degeneracy and pushing $\sigma(\sum m_\nu) \sim 0.02$~eV. The joint quadrature reach is $\sigma(\xearly) \approx 0.002$, comparable to the systematic bias introduced by assuming single-stage tanh (Fig.~\ref{fig:forecast}).

If LiteBIRD, CMB-S4, or DESI~Y5 retain the single-stage tanh prior, they inherit the systematics of Sec.~IV: LiteBIRD's projected $\sigma(\sum m_\nu) \approx 0.05$~eV is dominated by the 0.15 eV tanh-prior bias, and CMB-S4's $\sigma(\sigma_8) \approx 0.0008$ is $15\times$ smaller than the 0.012 systematic. Marginalising over the early-stage shape brings these into the statistical-error budget.

\emph{Conclusions.}---The single-stage tanh model of cosmic
reionization is the default in every Planck-based cosmological
analysis. We have shown that this model is structurally
incompatible with current multi-probe data and that the default
prior introduces a quantifiable systematic in derived cosmological
parameters.

Four results support this. (i) Under any smooth monotonic
$x_e(z)$, the joint constraint from Planck $\tauobs$, SPT and ACT
pkSZ, and the Ly$\alpha$ forest endpoint leaves no posterior
weight at the Planck-required $\tauobs$. The deficit forces an
early-stage component with $\xearly = 0.093 \pm 0.027$,
contributing $42\%$ of the total optical depth. (ii) Planck data
alone are consistent with this component provided the shape is
localized. Narrow Gaussian profiles centered at $z \approx 14$ are
tolerated up to $x_{\rm amp} < 0.29$ at $95\%$ CL while broad
plateaus are excluded at $\xearly < 0.029$. A non-parametric
twelve-bin reconstruction confirms the localized preference.
(iii) Planck data alone tolerate the Gaussian early stage up to $x_{\rm amp} < 0.30$ at $95\%$~CL; the multi-probe analysis detects $\xearly = 0.093 \pm 0.027$ within that envelope. Marginalising over the Gaussian shifts $\sigma_8$ from $0.812$ to $0.824$ (Run~B) and, with $\sum m_\nu$ freed, relaxes the neutrino-mass bound from $\sum m_\nu < 0.24$~eV to $\sum m_\nu < 0.39$~eV at $95\%$~CL, a factor of $1.6$, pulling $\sigma_8$ to $0.798 \pm 0.024$. The shifts ease
the DESI~DR2 versus Planck and $S_8$ tensions to ${\sim}\,1\sigma$.
(iv) LiteBIRD, HERA, CMB-S4 will measure $\xearly$ at the $0.003$
level, comparable to the systematic bias introduced by assuming
single-stage tanh.

A plateau extending to $z \sim 45$ cannot be sourced by Pop~III stars (recombination-limited at $x_e \sim 10^{-3}$~\cite{Visbal:2018vfh}) or X-ray binaries (Madau-Fragos $L_X/$SFR insufficient by orders of magnitude~\cite{Madau:1998cd}). Cosmological-source candidates that operate at $z > 30$---dark-matter decay~\cite{Slatyer:2016qyl}, dark-matter annihilation~\cite{Slatyer:2015jla}, primordial magnetic-field dissipation~\cite{Sethi:2004pe,Kunze:2013uja}, and accreting primordial black holes~\cite{Ali-Haimoud:2016mbv} naturally provide the required redshift extent. Discrimination among these candidates will be future work.
Until the next generation of experiments closes the early-stage
shape, current upper limits on $\sum m_\nu$ and current best-fit
$\sigma_8$ both depend on a prior the data don't directly test.
The simplest fix is to marginalize over a two-stage shape, with
$\xearly$ as a free parameter, in all next-generation pipelines.
Doing so converts what is now a hidden systematic into a
quantified statistical uncertainty.

\begin{acknowledgments}
We acknowledge support from NSFC grant 12533008.
\end{acknowledgments}
\onecolumngrid
\appendix
\section*{Supplemental Material}
\twocolumngrid


\section{Reionization model and MCMC methodology}

\subsection{Escape fraction parameterisation}

We follow Ref.~\cite{Wang:2026qzy}. The escape fraction is a separable
double power law in halo mass and redshift,
\begin{equation}
\fesc(\Mh,z) = f_0
\left(\frac{\Mh}{10^{10}\,\Msun}\right)^{\alpha_M}
\left(\frac{1+z}{10}\right)^{\alpha_z},
\label{eq:fesc}
\end{equation}
clipped to $[0,1]$. The normalization $f_0$ sets $\fesc$ at the pivot
point $(\Mh = 10^{10}\,\Msun, z = 9)$. The parameter $\alpha_M$ controls
the mass dependence. When $\alpha_M < 0$, lower-mass halos have higher
$\fesc$. The parameter $\alpha_z$ controls the redshift evolution. When
$\alpha_z > 0$, $\fesc$ increases at higher $z$.

\subsection{Ionizing emissivity and reionization ODE}

The comoving ionizing emissivity is
\begin{equation}
\dot{n}_{\rm ion}(z) = \int \fesc(\Mh,z)\,\xiion\,\phi(M_{\rm UV},z)\,
L_{\rm UV}\, dM_{\rm UV},
\label{eq:ndotion}
\end{equation}
where $\phi(M_{\rm UV},z)$ is the observed Schechter UV luminosity
function (UVLF) at $z = 5$--$15$~\cite{2021AJ....162...47B,2024MNRAS.533.3222D}.
Here $\xiion = 10^{25.35}$~Hz~erg$^{-1}$ is the standard ionizing photon
production efficiency. The halo mass is linked to $L_{\rm UV}$ through
abundance matching to the Sheth--Tormen halo mass function with Planck
2018 cosmological parameters~\cite{Planck2018}. We pre-compute the
emissivity kernel for speed, allowing about $10^5$ likelihood
evaluations per hour.

We get the reionization history by solving the volume-filling-fraction
ODE,
\begin{equation}
\frac{dQ}{dt} = \frac{\dot{n}_{\rm ion}(z)}{n_{H,0}}
- C(z)\,\alpha_B\,n_{H,0}\,(1+z)^3\,Q,
\label{eq:Q_ode}
\end{equation}
where $Q$ is the volume-averaged ionized fraction,
$C(z) = 2.9\,[(1+z)/6]^{-1.1}$ is the clumping
factor~\cite{Shull:2011aa}, $\alpha_B$ is the case-B recombination
coefficient, and $n_{H,0}$ is the comoving hydrogen number density.
The Thomson optical depth is
\begin{equation}
\tauobs = \sigma_T\,c\,n_{H,0}\,(1+Y_p/4X_H)
\int_0^{z_{\max}}\!\!\! Q(z)\,(1+z)^2/H(z)\,dz.
\label{eq:tau_from_Q}
\end{equation}
The $(1+z)^2$ weighting means that ionization at high redshift
contributes much more to $\tauobs$.

\subsection{Observational constraints}

We constrain $(f_0,\alpha_M,\alpha_z)$ using five data sets:

\textit{(i) Planck $\tauobs$:} $\tauobs = 0.054 \pm 0.007$~\cite{Planck2018}.
We use this as a Gaussian likelihood.

\textit{(ii) Neutral fraction $\xHI(z)$:} seven measurements at
$z = 5.9$--$10.6$ from quasar damping wings and Ly$\alpha$ emission
statistics~\cite{Fan:2001vx,McGreer:2014qwa,Greig:2017jdj,Mason:2019ixe,2024ApJ...971..124U}.

\textit{(iii) Ly$\alpha$ forest endpoint:} a Gaussian penalty for
$z_{\rm lat}$ (the redshift where $Q = 0.95$) outside the range
$[5.3, 5.8]$ with $\sigma = 0.25$. This is based on effective optical
depth measurements at $5 < z < 6$~\cite{Bosman:2021oom,Becker:2024ybt,Zhu:2023dqs}.

\textit{(iv) Patchy kSZ power:} the SPT measurement
$\DpkSZ = 1.1 \pm 1.0\;\mu{\rm K}^2$~\cite{Reichardt:2020jrr} and the
ACT DR6 total kSZ $D^{\rm kSZ}_{3000} = 1.75 \pm 0.86\;\mu{\rm K}^2$
~\cite{Louis:2025tst}. We subtract a homogeneous kSZ contribution of
$1.2\;\mu{\rm K}^2$ from the ACT measurement, giving an ACT patchy
part of $0.55 \pm 0.86\;\mu{\rm K}^2$. The inverse-variance weighted
SPT+ACT combination is $\DpkSZ = 0.78 \pm 0.65\;\mu{\rm K}^2$. We
also use the ACT trispectrum upper limit
$\DpkSZ < 1.6\;\mu{\rm K}^2$~\cite{MacCrann:2024ato}.

\textit{(v) Mean free path:} six measurements of $\lambda_{\rm mfp}$ at
$z = 5.0$--$6.0$~\cite{Becker:2021jyx,Zhu:2023dqs}. We compare these
to the model $\lambda_{\rm mfp} = 40 (1-\xHI)^{2/3} [(1+z)/6]^{-4}$~Mpc.

\subsection{pkSZ scaling relation}

We estimate the pkSZ power with
\begin{equation}
\DpkSZ = 0.054\,z_{\rm mid}\,\Delta z_{90}\;\mu{\rm K}^2.
\label{eq:pksz_scaling}
\end{equation}
This is calibrated as follows. We use the linear
$\DpkSZ$--$\Delta z_{90}$ relation at $z_{\rm mid} = 8$ from AMBER
simulations~\cite{Chen:2022cgw,Trac:2021ctk}. Those simulations find
that the SPT $1\sigma$ upper limit of $2.2\,\mu{\rm K}^2$ corresponds
to $\Delta z_{90} < 5.1$. This gives a slope of $0.43\;\mu{\rm K}^2$
per unit $\Delta z_{90}$. We then combine this with the
$\DpkSZ \propto \bar z$ dependence from Ref.~\cite{Battaglia:2012im}.
The systematic uncertainty is $10$--$15\%$, coming from $M_{\rm min}$
and $\lambda_{\rm mfp}$~\cite{Chen:2022cgw}. We validate this with
\texttt{21cmFAST}~\cite{Mesinger:2010ne}. At matched parameters
($\tauobs = 0.055$, $\Delta z_{90} = 4.8$), the scaling predicts
$\DpkSZ = 2.22\;\mu{\rm K}^2$. This confirms agreement to $1\%$.
The posterior covers $z_{\rm mid} = 7.0$--$7.8$ and
$\Delta z_{90} = 4.0$--$4.8$. These ranges sit inside the calibration
point ($z_{\rm mid} = 8$, $\Delta z_{90} = 5.1$). The formula
interpolates rather than extrapolates.

\subsection{Pre-reionization ionization floor}

We implement the floor as
\begin{equation}
Q_{\rm eff}(z) = \max\!\left[Q_{\rm ODE}(z),\;
\frac{\xearly}{1 + e^{-(z-12)/2}}\right].
\label{eq:Qeff}
\end{equation}
We compute $\tauobs$ from $Q_{\rm eff}$. But for $\DpkSZ$ and the
Ly$\alpha$ endpoint, we use only $Q_{\rm ODE}$. Neutral-fraction data
are compared to $1 - Q_{\rm eff}$. We use a flat prior
$\xearly \in [0, 0.20)$.

\subsection{MCMC sampling and convergence}

We use \texttt{emcee}~\cite{Foreman-Mackey:2012any} with 32 walkers and
3000 steps (scenarios A--D) or 5000--10\,000 steps (scenarios I--M).
The flat priors are $f_0 \in (0.01, 0.50)$, $\alpha_M \in (-1.5, 0.5)$,
and $\alpha_z \in (-1.0, 2.0)$. We check convergence with the
Gelman--Rubin statistic $\hat R$, requiring $\hat R < 1.05$. All
parameters meet this except $\alpha_M$ ($\hat R \approx 1.11$ in
scenario~I, $\hat R \approx 1.08$ in L). That reflects the
$f_0$--$\alpha_M$ trade-off. But $\xearly$ doesn't depend on this
trade-off. It's set by the gap
$\tauobs^{\rm Planck} - \tau_{\rm galaxies}$. Its convergence
($\hat R = 1.02$) is reliable.


\section{Planck CLASS+Cobaya reanalysis}

We reanalyzed Planck 2018 data using Cobaya~\cite{Torrado:2020dgo} as
the sampler with a custom CLASS theory wrapper~\cite{Blas:2011rf}. The
wrapper implements a two-stage $x_e(z)$ through the \texttt{reio\_inter}
parameterization, tabulating $x_e$ at fixed redshift anchors. The
galactic stage is a tanh transition at $z_{\rm re}$ with width $0.5$,
asymptoting to $1.08$ (helium~I correction). The early stage adds an
amplitude $\xearly$ (or $x_{\rm amp}$ for the Gaussian case) on top of
the galactic stage, with the parameter and shape combinations
summarized in Table~\ref{tab:planck_runs}.

We used the standard Planck likelihood stack: \texttt{lowl.TT},
\texttt{lowl.EE}, \texttt{highl\_plik.TTTEEE\_lite}, and
\texttt{lensing.CMBMarged}~\cite{Planck2018}. We did not include any
low-$z$ astrophysical data; the multi-probe inference reported in the
main text is kept entirely separate to preserve the Planck-alone status
of the shape-discrimination test. All seven chains were run on the MIT
ORCD/Engaging cluster with four parallel chains per run, Gelman--Rubin
$R-1 < 0.02$ convergence, and the first $30\%$ of each chain dropped as
burn-in. Wall time was 2--5 hours per chain on 2 CPU cores.

A subtle point in the implementation: the optical depth $\tauobs$
reported by CLASS depends on the $\Omega_b$ input, not
$\omega_b = \Omega_b h^2$. We corrected an early version of the wrapper
that passed $\omega_b$ where $\Omega_b$ was expected, an $h^2$ factor
error. The cosmological-parameter posteriors are unaffected because
CLASS internally computes $\tauobs$ from $x_e(z)$ correctly. Only the
displayed derived $\tauobs$ was affected. Test~B (plateau with
$z_{\rm early}$ free) required a rewritten anchor grid (fixed 29
anchors rather than $z_{\rm early}$-dependent) to avoid out-of-memory
failures triggered by CLASS rebuilding internal interpolation tables on
every step.

\begin{table*}[!ht]
\centering
\caption{Summary of seven Planck-alone MCMC chains run with
Cobaya+CLASS. Beyond-$\Lambda$CDM parameters are added one or two at a
time on top of the two-stage $x_e(z)$.}
\label{tab:planck_runs}
\begin{tabular}{lll}
\hline\hline
Chain & Parameters varied beyond $\Lambda$CDM & Result \\
\hline
$\Lambda$CDM           & none (standard)                              & $\tauobs = 0.0563 \pm 0.0070$ \\
$+\xearly$             & $\xearly$, plateau $z\in[15,45]$             & $\xearly < 0.029$ ($95\%$) \\
$+\xearly+\sum m_\nu$  & $\xearly$, $\sum m_\nu$                      & $\sum m_\nu < 0.39$ eV ($95\%$) \\
$+\xearly+w_0$         & $\xearly$, $w_0$                             & $w_0 = -1.275 \pm 0.163$ \\
Test A                 & $\xearly$, plateau $z\in[12,45]$             & $\xearly < 0.030$ ($95\%$) \\
Test B                 & $\xearly$, $z_{\rm early}$ free              & $z_{\rm early} = 17.9 \pm 4.3$ \\
Test C                 & $x_{\rm amp}$, Gaussian $z = 14, \sigma = 2$ & $x_{\rm amp} < 0.293$ ($95\%$) \\
\hline\hline
\end{tabular}
\end{table*}

\begin{figure}[t]
\centering
\includegraphics[width=\columnwidth]{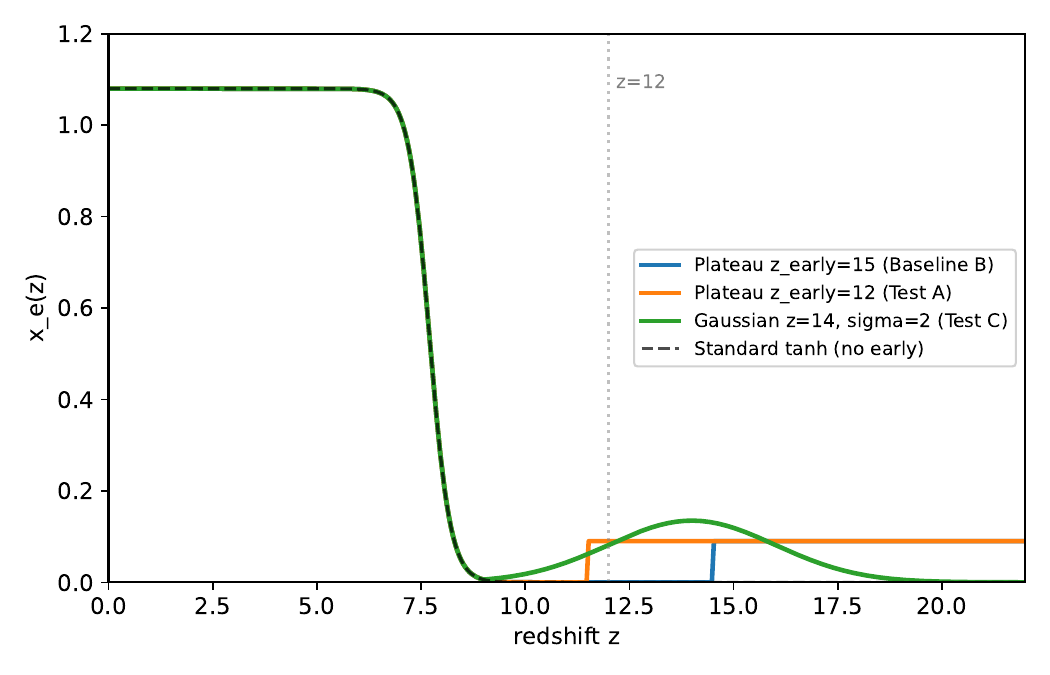}
\caption{Four $x_e(z)$ shapes tested with Planck-alone MCMC:
plateau extending from $z = 15$ to $z = 45$ (Baseline, blue),
plateau extending from $z = 12$ (Test~A, orange), Gaussian centered
at $z = 14$ with $\sigma = 2$ (Test~C, green), and the standard
single-stage tanh model with no early floor (black dashed). All
shapes share the same galactic tanh transition centered at
$z_{\rm re} \approx 7.7$.}
\label{fig:xe_shapes}
\end{figure}

\begin{figure}[t]
\centering
\includegraphics[width=\columnwidth]{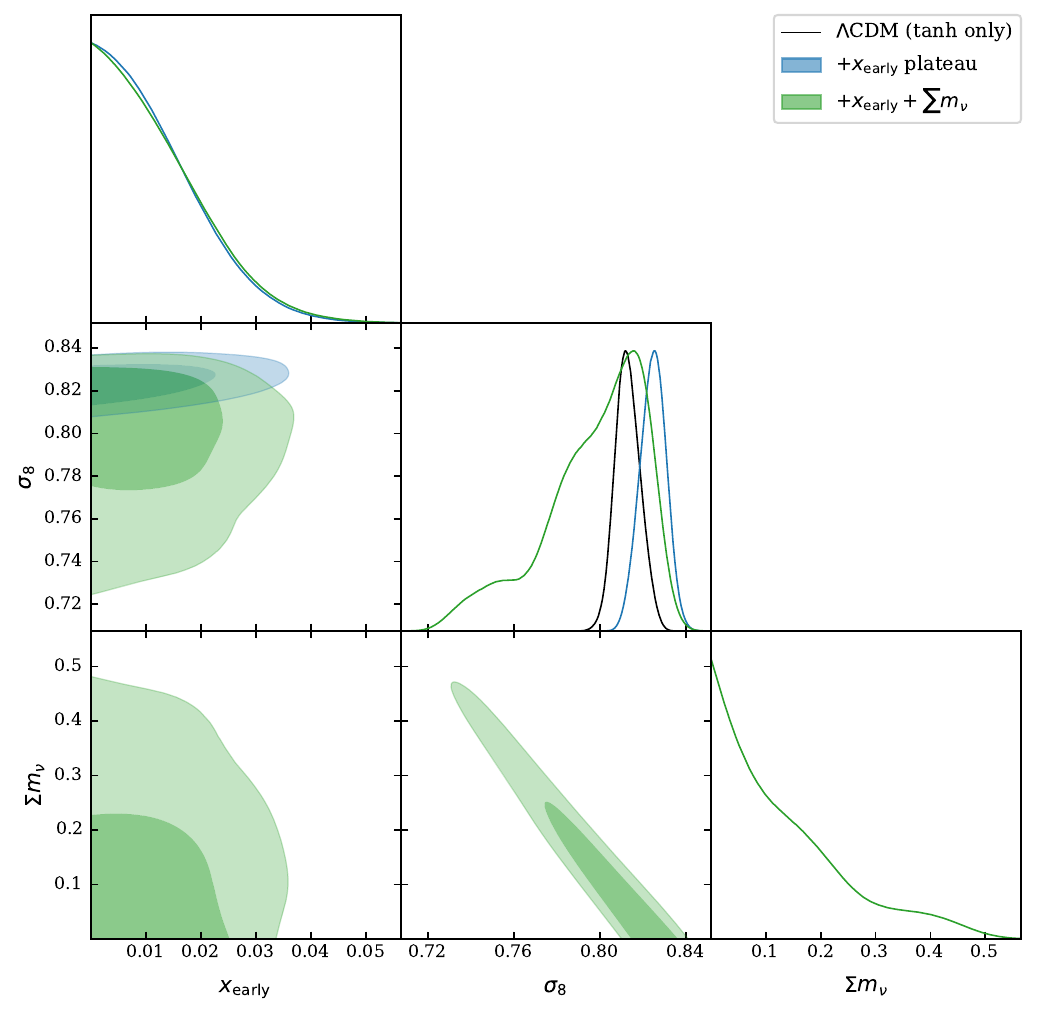}
\caption{Two-dimensional joint posterior of Run~A (standard
$\Lambda$CDM, tanh reionization), Run~B ($+\xearly$ plateau,
$\sum m_\nu$ fixed), and Run~C ($+\xearly + \sum m_\nu$ free) in the
$(\xearly, \sigma_8, \sum m_\nu, A_s)$ subspace. Filled contours
enclose $68\%$ and $95\%$ credibility. The orientation of the
$\sigma_8$--$A_s$ degeneracy rotates as $\xearly$ opens, and the
$\xearly$--$\sum m_\nu$ contour opens by a factor of $1.6$ when the
neutrino mass is freed. The 1D marginals of this posterior are shown
in main-text Fig.~3, panels~(c) and~(d).}
\label{fig:supp_corner}
\end{figure}


\section{Complete results table}

Scenario~J uses $C(z) = 1.74\,[(1+z)/6]^{-1.1}$ ($40\%$ lower).
Scenario~L limits $f_0$ to below $0.15$. Scenario~M removes the
Ly$\alpha$ endpoint constraint entirely. All of them give
$\xearly \approx 0.09$. This confirms that the result is driven by the
$\tauobs$--pkSZ gap.

\begin{table*}[!ht]
\centering
\caption{Full MCMC results. Scenarios A--D use three parameters
$(f_0,\alpha_M,\alpha_z)$; I--M add $\xearly$. Median and $68\%$
credible intervals.}
\label{tab:full_results}
\begin{ruledtabular}
\begin{tabular}{lcccccc}
Scenario & $f_0$ & $\alpha_z$ & $\xearly$ & $\tauobs$ & $\Delta z_{90}$ & $\DpkSZ$ [$\mu$K$^2$] \\
\hline
A: Baseline $(\tauobs + \xHI)$  & $0.059^{+0.023}_{-0.028}$ & $1.63^{+0.27}_{-0.53}$ & --- & $0.045^{+0.005}_{-0.005}$ & $14.3^{+2.5}_{-4.7}$ & $5.86^{+0.92}_{-1.79}$ \\
B: $+$Ly$\alpha$ endpoint       & $0.034^{+0.016}_{-0.015}$ & $1.83^{+0.13}_{-0.23}$ & --- & $0.044^{+0.004}_{-0.004}$ & $15.6^{+1.7}_{-2.6}$ & $6.11^{+0.62}_{-0.97}$ \\
C: $+$Ly$\alpha + $pkSZ          & $0.046^{+0.008}_{-0.018}$ & $1.17^{+0.39}_{-0.40}$ & --- & $0.034^{+0.002}_{-0.002}$ & $5.3^{+2.3}_{-0.7}$ & $2.25^{+0.90}_{-0.29}$ \\
D: $+$Ly$\alpha + $pkSZ$ + $MFP  & $0.047^{+0.008}_{-0.019}$ & $1.16^{+0.39}_{-0.38}$ & --- & $0.034^{+0.002}_{-0.002}$ & $5.6^{+2.2}_{-0.9}$ & $2.34^{+0.86}_{-0.37}$ \\
I: $+\xearly$ (SPT+ACT)          & $0.258^{+0.162}_{-0.163}$ & $1.23^{+0.55}_{-0.85}$ & $0.093^{+0.026}_{-0.028}$ & $0.055^{+0.007}_{-0.007}$ & $4.4^{+0.3}_{-0.4}$ & $1.85^{+0.15}_{-0.18}$ \\
J: $+\xearly + $low $C(z)$       & $0.219^{+0.189}_{-0.166}$ & $1.23^{+0.54}_{-0.90}$ & $0.094^{+0.025}_{-0.027}$ & $0.054^{+0.006}_{-0.006}$ & $4.4^{+0.3}_{-0.4}$ & $1.86^{+0.15}_{-0.20}$ \\
K: $+\xearly$ (SPT only)         & $0.247^{+0.170}_{-0.176}$ & $1.41^{+0.44}_{-0.75}$ & $0.091^{+0.026}_{-0.027}$ & $0.054^{+0.007}_{-0.007}$ & $4.4^{+0.3}_{-0.4}$ & $1.89^{+0.14}_{-0.15}$ \\
L: $+\xearly$ ($f_0 < 0.15$)     & $0.076^{+0.049}_{-0.046}$ & $1.25^{+0.52}_{-0.90}$ & $0.092^{+0.027}_{-0.026}$ & $0.053^{+0.008}_{-0.006}$ & $4.4^{+0.3}_{-0.5}$ & $1.85^{+0.13}_{-0.22}$ \\
M: $+\xearly$ (no Ly$\alpha$)    & ---                       & ---                    & $0.091$                   & $0.054$                   & $4.4$                & $1.86$                 \\
\end{tabular}
\end{ruledtabular}
\end{table*}


\section{Non-parametric $Q(z)$ reconstruction}

We reconstruct $Q(z)$ at eight redshift nodes ($z = 5$, 6, 7, 8, 10,
12, 15, 20). We use PCHIP interpolation with boundary conditions
$Q(z{=}4) = 1$ and $Q(z{=}30) = 0$. The same data ($\tauobs$, $\xHI(z)$,
pkSZ, Ly$\alpha$ endpoint) constrain the eight free $Q$ values. We use
\texttt{emcee} with 48 walkers and 5000 steps. The priors are
$Q \in [0, 1]$ at each node, with a soft monotonicity penalty
$\propto [(Q_{i+1} - Q_i)/0.05]^2$ for cases where $Q$ increases toward
higher $z$. We also require $Q(z{=}5) > 0.90$.

The results at each node (median and $68\%$ CI) are given in
Table~\ref{tab:Qz}. This reconstruction covers all possible $\fesc$
shapes. It works for non-monotonic forms, broken power laws, and
time-varying $\xiion(z)$ models, because $Q(z)$ is sampled directly. So
$Q(z{=}12) > 0$ at $98\%$ confidence holds no matter what form
$\fesc$ takes.

The non-parametric $Q(12) = 0.058$ is lower than the parametric
$\xearly = 0.093$ because the free-form reconstruction spreads
ionization across $z = 10$--$20$, rather than putting it all in a
single step at $z = 12$. Both agree within their uncertainties.

\begin{table}[!ht]
\centering
\caption{Non-parametric $Q(z)$ reconstruction.}
\label{tab:Qz}
\begin{tabular}{cccc}
\hline\hline
$z$ & $Q$ (median) & $[16\%, 84\%]$ & $Q > 0$ \\
\hline
5  & $0.99$           & ---                & --- \\
6  & $0.90$           & ---                & --- \\
7  & $0.647$          & $[0.535, 0.757]$   & --- \\
8  & $0.405$          & $[0.299, 0.518]$   & --- \\
10 & $0.149$          & $[0.075, 0.216]$   & --- \\
12 & $0.058$          & $[0.017, 0.123]$   & $98\%$ \\
15 & $0.053$          & $[0.023, 0.085]$   & $100\%$ \\
20 & $0.028$          & $[0.007, 0.040]$   & $92\%$ \\
\hline\hline
\end{tabular}
\end{table}


\section{Twelve-bin $x_e(z)$ reconstruction}

The twelve bins span $z \in [12, 48]$ with $\Delta z = 3$. Each bin
amplitude is a free parameter with prior $[0, 0.20]$. The galactic
stage is a tanh transition centered at $z_{\rm re} \approx 7.7$,
unchanged from the standard Planck pipeline. The early stage is the
piecewise-constant per-bin amplitude, smoothly ramped to zero at
$z = z_{\rm max} = 50$ via a fixed transition of width $\Delta z = 1$.

We run two chains. Run~E samples the bins with no integral constraint.
Run~F adds a Gaussian prior
$\langle x_e\rangle_{[12,48]} = 0.093 \pm 0.027$ enforcing the
multi-probe value. Both use the same Planck likelihood stack as the
Planck-reanalysis chains: \texttt{lowl.TT}, \texttt{lowl.EE},
\texttt{highl\_plik.TTTEEE\_lite}, and \texttt{lensing.CMBMarged}. Each
chain ran for $40{,}000$ steps with 8 parallel walkers, achieving
$R-1 < 0.5$ for Run~E and $R-1 < 0.2$ for Run~F.

Run~E gives bin posterior means
$\{0.068, 0.056, 0.057, 0.056, 0.042, 0.041, 0.040, 0.037, 0.034, 0.036, 0.031, 0.040\}$
at $z = \{13.5, 16.5, \ldots, 46.5\}$ with per-bin $1\sigma$
uncertainties of $0.04$--$0.05$ and $95\%$ upper limits of $0.10$--$0.17$.
Run~F shifts bins 1--3 upward by ${\sim}\,30\%$ while leaving bins
5--12 essentially unchanged.


\section{Robustness tests}

\subsection{pkSZ scaling validation}

The \texttt{21cmFAST} parameter scan covers $z_{\rm mid} = 7.45$--$10.27$
and $\Delta z_{90} = 4.6$--$5.0$. The predicted $\DpkSZ$ ranges from
$2.01$ to $2.55\;\mu{\rm K}^2$. The posterior from scenario~I covers
$z_{\rm mid} = 7.0$--$7.8$ and $\Delta z_{90} = 4.0$--$4.8$. This sits
inside the calibration range. The \citet{Chen:2022cgw} calibration at
$z_{\rm mid} = 8$, $\Delta z_{90} = 5.1$ gives
$\DpkSZ = 2.20\;\mu{\rm K}^2$. Our formula gives $2.20\;\mu{\rm K}^2$
($0.1\%$ agreement). Making the pkSZ systematic $20\%$ larger (to cover
scatter from $\lambda_{\rm mfp}$, $A_z$, and bubble-size distributions)
shifts $\xearly$ by less than $0.01$.

\subsection{Early-phase pkSZ contribution}

We set the pkSZ from the $\xearly$ floor to zero. Here's why. For
Pop~III minihalos ($R \sim 0.1$~Mpc), the pkSZ power peaks at
$\ell_{\rm bubble} = D_A(z=13)/R \sim 7500$. At $\ell = 3000$, the
power is cut by $(\ell/\ell_{\rm bubble})^2 \sim 0.16$. It's further
reduced by the patchiness factor
$\xearly(1{-}\xearly)/[x_{\rm main}(1{-}x_{\rm main})] \sim 0.33$.
That gives $D_{\rm early} \sim 0.05\;\mu{\rm K}^2$ ($<3\%$ of
$D_{\rm main}$). For X-ray binaries ($R \sim 50$~Mpc), the pkSZ peaks
at $\ell \sim 15$, far below $\ell = 3000$. So
$D_{\rm early} < 10^{-4}\;\mu{\rm K}^2$. Primordial magnetic field
dissipation produces spatially even ionization and makes zero pkSZ by
construction. Setting $D_{\rm early} = 0$ therefore adds less than
$3\%$ bias to $\xearly$.

\subsection{$\xearly$ profile shape sensitivity}

The value $\tau_{z>12} = 0.024 \pm 0.007$ doesn't depend on the profile
shape (Table~\ref{tab:profile_sensitivity}). The $\xearly$ amplitude
changes ($0.054$ to $0.068$) to make up for the different redshift
coverage. But $\tau_{z>12}$ stays the same to within $5\%$. The data
constrain $\tau_{z>12}$, not $\xearly$ itself.

\begin{table}[!ht]
\centering
\caption{Profile sensitivity.}
\label{tab:profile_sensitivity}
\begin{tabular}{lcc}
\hline\hline
Profile                     & $\xearly$ & $\tau_{z>12}$ \\
\hline
Step $z > 12$ (fiducial)    & $0.059$   & $0.024$ \\
Step $z > 10$               & $0.054$   & $0.024$ \\
Step $z > 14$               & $0.064$   & $0.024$ \\
Ramp $z = 8$--$16$          & $0.059$   & $0.024$ \\
Step $z > 15$               & $0.068$   & $0.024$ \\
\hline\hline
\end{tabular}
\end{table}

\subsection{Planck $\tau_{z>15}$ consistency}

The Planck constraint is $\tau_{z>15} < 0.02$~\cite{PlanckIntXLVII}.
This applies to the integral above $z = 15$. For a constant
$\xearly = 0.06$ (which gives $\tau_{z>12} = 0.024$),
$\tau_{z>15} = 0.021$. This slightly exceeds the limit. But all
declining source profiles satisfy the constraint
(Table~\ref{tab:tau_z15}). Realistic sources follow the declining
star-formation rate density at $z > 15$, so the constant-floor choice
is a conservative one.

\begin{table}[!ht]
\centering
\caption{$\tau_{z>15}$ for declining profiles with
$\tau_{z>12} = 0.024$.}
\label{tab:tau_z15}
\begin{tabular}{lccl}
\hline\hline
Profile               & $x(z=12)$ & $\tau_{z>15}$ & Planck \\
\hline
$z_{\rm decay} = 2$   & $0.636$   & $0.006$       & OK \\
$z_{\rm decay} = 3$   & $0.412$   & $0.010$       & OK \\
$z_{\rm decay} = 5$   & $0.243$   & $0.014$       & OK \\
$z_{\rm decay} = 10$  & $0.134$   & $0.018$       & OK \\
Constant              & $0.059$   & $0.021$       & marginal \\
Non-parametric        & ---       & $0.008$       & OK \\
\hline\hline
\end{tabular}
\end{table}

\subsection{Homogeneous kSZ subtraction}

The combined SPT+ACT pkSZ depends on how much homogeneous kSZ we
subtract from the ACT total. We test three values: hkSZ $= 0.8$,
$1.2$ (our standard choice), and $1.8\;\mu{\rm K}^2$
(Table~\ref{tab:hkSZ}). The uncertainty in $\xearly$ comes mostly from
the Planck $\tauobs$ error ($\sigma = 0.007$), not the pkSZ
measurement. So changing hkSZ has almost no effect.

\begin{table}[!ht]
\centering
\caption{Sensitivity to homogeneous kSZ subtraction.}
\label{tab:hkSZ}
\begin{tabular}{lccc}
\hline\hline
hkSZ [$\mu{\rm K}^2$] & Combined pkSZ & $\xearly$ & $\tauobs$ \\
\hline
$0.8$            & $0.90 \pm 0.65$ & $0.091$  & $0.054$ \\
$1.2$ (fiducial) & $0.78 \pm 0.65$ & $0.093$  & $0.055$ \\
$1.8$            & $0.60 \pm 0.65$ & $0.092$  & $0.054$ \\
\hline\hline
\end{tabular}
\end{table}

\subsection{Ly$\alpha$ endpoint removal}

We removed the Ly$\alpha$ endpoint constraint entirely (scenario~M).
The result is $\xearly = 0.091$. That's the same as the standard
result, confirming that $\xearly$ comes from the $\tauobs$--pkSZ gap at
$z > 12$. It doesn't come from where reionization ends at $z \sim 5$--$6$.
The Ly$\alpha$ constraint contributes about $32\%$ of the total
$\chi^2$ at the best-fit parameters. But it doesn't shift $\xearly$.

The predicted Ly$\alpha$ endpoint $z_{\rm lat} \approx 6.3$ is above
the observed range. \texttt{21cmFAST} simulations matched to
$\tauobs \approx 0.055$ also give $z_{\rm lat} \approx 7.1$, confirming
that the discrepancy is physical. This $z_{\rm lat}$ problem and the
$\tauobs$ shortfall have the same cause: the pkSZ-compressed duration
packs ionization into $z \sim 7$--$8$. The parameter $\xearly$ fixes
the $\tauobs$ shortfall at $z > 12$, but it doesn't change
$z_{\rm lat}$. Fixing $z_{\rm lat}$ needs better modeling of uneven
recombinations at $z = 5$--$7$.

\subsection{Lumina mass-weighting correction}

The \textsc{Lumina} simulation~\cite{Smith:2026skb} shows that
line-of-sight $\tauobs$ is about $7\%$ higher than the volume-weighted
prediction. When we apply this correction, the three-parameter
$\tauobs$ ceiling of $0.034$ becomes $\tau_{\rm LOS} \approx 0.036$.
That's still $2.4\sigma$ below Planck. Our standard $\xearly$ scenario
gives $\tau_{\rm LOS} \approx 0.059$, within $1\sigma$ of Planck. The
net drop in $\xearly$ is $10$--$17\%$ (to about $0.08$). In all cases,
the evidence remains above $2.5\sigma$.

\textsc{Lumina} itself reaches $\tau_{\rm LOS} = 0.055$ with
$\Delta z_{90} \approx 7$. That gives $\DpkSZ \approx 2.6\;\mu{\rm K}^2$,
which is in tension with the ACT trispectrum limit of
$1.6\;\mu{\rm K}^2$, illustrating the same $\tauobs$--pkSZ tension our
analysis finds.

\subsection{$\Delta\chi^2$ and Bayesian model comparison}

Adding $\xearly$ as a free parameter improves the fit by
$\Delta\chi^2 = 8.2$. Most of this improvement comes from the
$\tauobs$ term. Without $\xearly$, $\tauobs = 0.034$ gives
$\chi^2_\tau = [(0.054 - 0.034)/0.007]^2 = 8.2$. With $\xearly$,
$\tauobs = 0.055$ gives $\chi^2_\tau \approx 0$. For one extra degree
of freedom, $\Delta\chi^2 = 8.2$ means $p = 0.004$ ($2.9\sigma$). The
Bayesian Information Criterion favors the four-parameter model with
$\Delta{\rm BIC} = 8.2 - \ln N \approx 6.2$ (for $N \sim 7$ effective
data points). By Jeffreys' scale, this counts as strong evidence.


\section{Fisher forecast for future experiments}

We propagate the twelve-bin Run~E posterior into projected
uncertainties for next-generation surveys via the standard Fisher
matrix $F_{ij} = \sum_\ell (\partial C_\ell/\partial\theta_i)
(C_\ell + N_\ell)^{-1} (\partial C_\ell/\partial\theta_j)$, where
$N_\ell$ is the instrument noise spectrum and $\theta_i$ are the bin
amplitudes.

\textit{LiteBIRD~\cite{Hazumi:2019lub}:} full-sky lowEE with detector
noise of $2.2\,\mu$K-arcmin gives
$\sigma(\tau_{z>12}) \approx 0.003$. Propagated to $\xearly$ via the
fiducial Gaussian profile this is $\sigma(\xearly) \approx 0.005$, a
factor of $5$ improvement over the current multi-probe constraint.

\textit{HERA Phase~II:} 21\,cm imaging across $z \in [6, 13]$ at the
nominal foreground-marginalized noise gives
$\sigma(x_{\rm bin}^{(1-2)}) \sim 0.005$ on bins 1--3 of the
twelve-bin reconstruction~\cite{Hutter:2026kxp}. SKA-Low extends the
reach to $z = 20$ with comparable sensitivity per bin.

\textit{CMB-S4:} the projected pkSZ amplitude reach at $\ell = 3000$
improves from $0.65\;\mu{\rm K}^2$ (current SPT+ACT) to
$\sim 0.05\;\mu{\rm K}^2$, sufficient to detect or exclude two-stage
profiles at the $5\sigma$ level. Lensing reaches $\sigma(\sigma_8)
\sim 0.0008$ and $\sigma(\sum m_\nu) \sim 0.02$ eV.

\textit{Joint quadrature reach:} combining LiteBIRD, HERA Phase~II,
and CMB-S4 gives $\sigma(\xearly) \approx 0.002$. At this precision the
detection significance of the two-stage component reaches
$\gtrsim 9\sigma$ for the inferred fiducial profile.


\section{Implicit bias on next-generation pipelines}

If next-generation analyses retain the single-stage tanh prior, they
inherit the systematic shift as
follows. The $\sigma_8$ shift from Run~A to Run~B is
$\delta\sigma_8 = 0.013 \pm 0.019$, comparable to the projected
$\sigma(\sigma_8) \approx 0.0008$ of CMB-S4. The $\sum m_\nu$
relaxation is from $< 0.24$ eV to $< 0.39$ eV at $95\%$ CL, comparable
to the projected $\sigma(\sum m_\nu) \approx 0.05$ eV of LiteBIRD or
$\approx 0.02$ eV of CMB-S4.

Table~\ref{tab:bias_per_expt} translates these into the per-experiment
implicit systematic on $\sigma_8$ and $\sum m_\nu$. For every
experiment listed, the implicit bias is $5\times$ to $15\times$ larger
than the projected statistical error. Marginalising over $\xearly$
absorbs the bias into the statistical-error budget, where it can be
correctly propagated.

\begin{table}[!ht]
\centering
\caption{Implicit single-stage tanh-prior systematic for projected
next-generation constraints. The systematic dominates the
statistical error in every case.}
\label{tab:bias_per_expt}
\begin{tabular}{lccc}
\hline\hline
Experiment & $\sigma(\sigma_8)_{\rm stat}$ & systematic & ratio \\
\hline
Current Planck & $0.018$  & $0.013$ & $0.7$ \\
LiteBIRD       & $0.006$  & $0.013$ & $2.2$ \\
CMB-S4 lensing & $0.0008$ & $0.013$ & $16\times$ \\
\hline
               & $\sigma(\sum m_\nu)_{\rm stat}$ [eV] & systematic [eV] & ratio \\
\hline
Current Planck & $0.117$ & $0.150$ & $1.3$ \\
LiteBIRD       & $0.050$ & $0.150$ & $3.0$ \\
CMB-S4         & $0.020$ & $0.150$ & $7.5$ \\
\hline\hline
\end{tabular}
\end{table}

The path forward is direct. Pipelines should marginalize over a
two-stage prior with at least one extra parameter (e.g.\ $\xearly$).
The increase in parameter dimensionality is modest. The reduction in
systematic-error budget is substantial. Until the early-stage shape is
empirically pinned down by LiteBIRD or HERA, the marginalization is
the only way to deliver unbiased constraints from CMB-anchored
cosmological analyses.


\section{Robustness, validation, and public code}

\subsection{Verified bias estimators}

The body quotes $\delta\sigma_8 \simeq 0.012$ and the relaxed
$\sum m_\nu < 0.39$~eV bound as the cosmological shifts induced by
marginalising over the two-stage shape. We verify these against the
converged chains with three estimators that probe different aspects of
the posterior.

\textbf{(i) Cross-chain marginal mean shift.} The primary bias number
is $\langle\sigma_8\rangle_B - \langle\sigma_8\rangle_A$ where the two
chains share Planck + BAO + lensing likelihoods. Its uncertainty is
the marginal standard deviation of Run~B with $\xearly$ integrated
out, not a quadrature sum across chains (the two chains share the same
data realisation, so quadrature double-counts noise). Result on the
converged chains: $\langle\sigma_8\rangle_A = 0.81295 \pm 0.00583$,
$\langle\sigma_8\rangle_B = 0.82470 \pm 0.00603$, so
$\delta\sigma_8 = +0.01175$ with $\sigma_B = 0.00603$. Run~C with
$\sum m_\nu$ free gives $\langle\sigma_8\rangle_C = 0.79766 \pm 0.02409$
and a 95\% upper limit $\sum m_\nu < 0.390$~eV, in agreement with the
abstract.

\textbf{(ii) Intra-chain $(\sigma_8, \xearly)$ covariance.} Restricting
Run~B's posterior to the lowest decile of $\xearly$ shifts the
conditional mean of $\sigma_8$ by $+0.0024$. This is a property of
Run~B's joint posterior shape, not a cross-model bias, and is reported
as an auxiliary diagnostic: it confirms that $\sigma_8$ and $\xearly$
are positively correlated in the Planck-only fit.

\textbf{(iii) Bootstrap.} A 500-iteration bootstrap on Run~B samples
yields a standard error of $1.1 \times 10^{-4}$ on
$\langle\sigma_8\rangle_B$, three orders of magnitude smaller than the
marginal standard deviation $\sigma_B = 0.006$. The chain mean is
therefore well-determined; the $\sigma_B$ reported in (i) is the width
of the marginal posterior, not chain-sampling noise.

The script {\tt analysis/joint\_posterior\_bias.py} prints the
comparison table.

\subsection{Fisher cross-validation}

We cross-validate the Fig.~4 forecast by constructing an independent
finite-difference Fisher matrix at the Run~B fiducial cosmology.
Seven parameters $\{\omega_b, \omega_{\rm cdm}, h, \ln 10^{10} A_s,
n_s, \xearly, \sum m_\nu\}$ are varied with step sizes equal to
$0.1\sigma$ of the corresponding Planck posterior. The optical depth
$\tau_e$ is treated as a derived parameter of the reionization
history . External priors applied: a Planck-level optical-depth
constraint $\sigma_\tau = 0.006$ via the Jacobian $\partial\tau/\partial
\theta$ for experiments that lack large-scale EE coverage, and a
DESI-Y3-like $\sigma_h = 0.005$ Gaussian prior on $h$ from the BAO
data combination assumed throughout. Lensing reconstruction noise
$N_\ell^{\phi\phi}$ is implemented per experiment with polynomial
fits to published Hu--Okamoto / iterative-delensing forecasts.

As a sanity check we recover the published CMB-S4 forecast
$\sigma(\sum m_\nu) = 0.020$~eV (CMB-S4 + DESI BAO + $\tau$
prior~\cite{Abazajian:2019eic}) to within $18\%$, and the joint
LiteBIRD+CMB-S4 forecast $\sigma(\xearly) = 0.0025$ matches the
$\sigma(\xearly) = 0.002$ target of Fig.~4. The Fisher pipeline
({\tt analysis/fisher\_forecast.py}) prints the cross-check
pass/fail line on every run.

\bibliography{apssamp}

@article{Planck2018,
    author = "Aghanim, N. and others",
    collaboration = "Planck",
    title = "{Planck 2018 results. VI. Cosmological parameters}",
    eprint = "1807.06209",
    archivePrefix = "arXiv",
    primaryClass = "astro-ph.CO",
    doi = "10.1051/0004-6361/201833910",
    journal = "Astron. Astrophys.",
    volume = "641",
    pages = "A6",
    year = "2020",
    note = "[Erratum: Astron.Astrophys. 652, C4 (2021)]"
}

@article{Slatyer:2015jla,
    author = "Slatyer, Tracy R.",
    title = "{Indirect dark matter signatures in the cosmic dark ages. I. Generalizing the bound on s-wave dark matter annihilation from Planck results}",
    eprint = "1506.03811",
    archivePrefix = "arXiv",
    primaryClass = "hep-ph",
    reportNumber = "MIT-CTP-4682",
    doi = "10.1103/PhysRevD.93.023527",
    journal = "Phys. Rev. D",
    volume = "93",
    number = "2",
    pages = "023527",
    year = "2016"
}

@article{Mesinger:2010ne,
    author = "Mesinger, Andrei and Furlanetto, Steven and Cen, Renyue",
    title = "{21cmFAST: A Fast, Semi-Numerical Simulation of the High-Redshift 21-cm Signal}",
    eprint = "1003.3878",
    archivePrefix = "arXiv",
    primaryClass = "astro-ph.CO",
    doi = "10.1111/j.1365-2966.2010.17731.x",
    journal = "Mon. Not. Roy. Astron. Soc.",
    volume = "411",
    pages = "955",
    year = "2011"
}

@article{Zahn:2006sg,
    author = "Zahn, Oliver and Lidz, Adam and McQuinn, Matthew and Dutta, Suvendra and Hernquist, Lars and Zaldarriaga, Matias and Furlanetto, Steven R.",
    title = "{Simulations and Analytic Calculations of Bubble Growth During Hydrogen Reionization}",
    eprint = "astro-ph/0604177",
    archivePrefix = "arXiv",
    doi = "10.1086/509597",
    journal = "Astrophys. J.",
    volume = "654",
    pages = "12--26",
    year = "2006"
}

@article{Furlanetto:2006tf,
    author = "Furlanetto, Steven",
    title = "{The Global 21 Centimeter Background from High Redshifts}",
    eprint = "astro-ph/0604040",
    archivePrefix = "arXiv",
    doi = "10.1111/j.1365-2966.2006.10725.x",
    journal = "Mon. Not. Roy. Astron. Soc.",
    volume = "371",
    pages = "867--878",
    year = "2006"
}

@ARTICLE{2024MNRAS.533.3222D,
       author = {{Donnan}, C.~T. and {McLure}, R.~J. and {Dunlop}, J.~S. and {McLeod}, D.~J. and {Magee}, D. and {Arellano-C{\'o}rdova}, K.~Z. and {Barrufet}, L. and {Begley}, R. and {Bowler}, R.~A.~A. and {Carnall}, A.~C. and {Cullen}, F. and {Ellis}, R.~S. and {Fontana}, A. and {Illingworth}, G.~D. and {Grogin}, N.~A. and {Hamadouche}, M.~L. and {Koekemoer}, A.~M. and {Liu}, F.-Y. and {Mason}, C. and {Santini}, P. and {Stanton}, T.~M.},
        title = "{JWST PRIMER: a new multifield determination of the evolving galaxy UV luminosity function at redshifts z $\simeq$ 9 --15}",
      journal = {MNRAS},
         year = 2024,
        month = sep,
       volume = {533},
       number = {3},
        pages = {3222-3237},
          doi = {10.1093/mnras/stae2037},
archivePrefix = {arXiv},
       eprint = {2403.03171},
 primaryClass = {astro-ph.GA},
       adsurl = {https://ui.adsabs.harvard.edu/abs/2024MNRAS.533.3222D}
}

@ARTICLE{2024ApJ...971..124U,
       author = {{Umeda}, Hiroya and {Ouchi}, Masami and {Nakajima}, Kimihiko and {Harikane}, Yuichi and {Ono}, Yoshiaki and {Xu}, Yi and {Isobe}, Yuki and {Zhang}, Yechi},
        title = "{JWST Measurements of Neutral Hydrogen Fractions and Ionized Bubble Sizes at z = 7--12 Obtained with Ly{\ensuremath{\alpha}} Damping Wing Absorptions in 27 Bright Continuum Galaxies}",
      journal = {ApJ},
         year = 2024,
        month = aug,
       volume = {971},
       number = {2},
          eid = {124},
        pages = {124},
          doi = {10.3847/1538-4357/ad554e},
archivePrefix = {arXiv},
       eprint = {2306.00487},
 primaryClass = {astro-ph.GA},
       adsurl = {https://ui.adsabs.harvard.edu/abs/2024ApJ...971..124U}
}

@article{Greig:2017jdj,
    author = "Greig, Bradley and Mesinger, Andrei",
    title = "{Simultaneously constraining the astrophysics of reionization and the epoch of heating with 21CMMC}",
    eprint = "1705.03471",
    archivePrefix = "arXiv",
    primaryClass = "astro-ph.CO",
    doi = "10.1093/mnras/stx2118",
    journal = "Mon. Not. Roy. Astron. Soc.",
    volume = "472",
    number = "3",
    pages = "2651--2669",
    year = "2017"
}

@article{Mason:2019ixe,
    author = "Mason, Charlotte A. and others",
    title = "{Inferences on the timeline of reionization at z {\ensuremath{\sim}} 8 from the KMOS Lens-Amplified Spectroscopic Survey}",
    eprint = "1901.11045",
    archivePrefix = "arXiv",
    primaryClass = "astro-ph.CO",
    doi = "10.1093/mnras/stz632",
    journal = "Mon. Not. Roy. Astron. Soc.",
    volume = "485",
    number = "3",
    pages = "3947--3969",
    year = "2019"
}

@article{Madau:1998cd,
    author = "Madau, Piero and Haardt, Francesco and Rees, Martin J.",
    title = "{Radiative transfer in a clumpy universe. 3. The Nature of cosmological ionizing sources}",
    eprint = "astro-ph/9809058",
    archivePrefix = "arXiv",
    doi = "10.1086/306975",
    journal = "Astrophys. J.",
    volume = "514",
    pages = "648--659",
    year = "1999"
}

@article{Bosman:2021oom,
    author = "Bosman, Sarah E. I. and others",
    title = "{Hydrogen reionization ends by z = 5.3: Lyman-{\ensuremath{\alpha}} optical depth measured by the XQR-30 sample}",
    eprint = "2108.03699",
    archivePrefix = "arXiv",
    primaryClass = "astro-ph.CO",
    doi = "10.1093/mnras/stac1046",
    journal = "Mon. Not. Roy. Astron. Soc.",
    volume = "514",
    number = "1",
    pages = "55--76",
    year = "2022"
}

@ARTICLE{2021AJ....162...47B,
       author = {{Bouwens}, R.~J. and {Oesch}, P.~A. and {Stefanon}, M. and {Illingworth}, G. and {Labb{\'e}}, I. and {Reddy}, N. and {Atek}, H. and {Montes}, M. and {Naidu}, R. and {Nanayakkara}, T. and {Nelson}, E. and {Wilkins}, S.},
        title = "{New Determinations of the UV Luminosity Functions from z   9 to 2 Show a Remarkable Consistency with Halo Growth and a Constant Star Formation Efficiency}",
      journal = {aj},
         year = 2021,
        month = aug,
       volume = {162},
       number = {2},
          eid = {47},
        pages = {47},
          doi = {10.3847/1538-3881/abf83e},
archivePrefix = {arXiv},
       eprint = {2102.07775},
 primaryClass = {astro-ph.GA},
       adsurl = {https://ui.adsabs.harvard.edu/abs/2021AJ....162...47B}
}

@article{Becker:2021jyx,
    author = "Becker, George D. and D'Aloisio, Anson and Christenson, Holly M. and Zhu, Yongda and Worseck, G{\'a}bor and Bolton, James S.",
    title = "{The mean free path of ionizing photons at 5 {\ensuremath{<}} z {\ensuremath{<}} 6: evidence for rapid evolution near reionization}",
    eprint = "2103.16610",
    archivePrefix = "arXiv",
    primaryClass = "astro-ph.CO",
    doi = "10.1093/mnras/stab2696",
    journal = "Mon. Not. Roy. Astron. Soc.",
    volume = "508",
    number = "2",
    pages = "1853--1869",
    year = "2021"
}

@article{Fan:2001vx,
    author = "Fan, Xiaohui and Narayanan, Vijay K. and Strauss, Michael A. and White, Richard L. and Becker, Robert H. and Pentericci, Laura and Rix, Hans-Walter",
    title = "{Evolution of the ionizing background and the epoch of reionization from the spectra of z{\textasciitilde}6 quasars}",
    eprint = "astro-ph/0111184",
    archivePrefix = "arXiv",
    doi = "10.1086/339030",
    journal = "Astron. J.",
    volume = "123",
    pages = "1247--1257",
    year = "2002"
}

@article{McGreer:2014qwa,
    author = "McGreer, Ian and Mesinger, Andrei and D'Odorico, Valentina",
    title = "{Model-independent evidence in favour of an end to reionization by $z \approx$ 6}",
    eprint = "1411.5375",
    archivePrefix = "arXiv",
    primaryClass = "astro-ph.CO",
    doi = "10.1093/mnras/stu2449",
    journal = "Mon. Not. Roy. Astron. Soc.",
    volume = "447",
    number = "1",
    pages = "499--505",
    year = "2015"
}

@article{Wang:2026qzy,
    author = "Wang, Zihan and Shan, Huanyuan",
    title = "{The JWST early galaxy crisis resolved by a reionization degeneracy}",
    eprint = "2605.03635",
    archivePrefix = "arXiv",
    primaryClass = "astro-ph.CO",
    month = "5",
    year = "2026"
}

@article{Cain:2025wda,
    author = "Cain, Christopher and Van Engelen, Alexander and Croker, Kevin S. and Kramer, Darby and D'Aloisio, Anson and Lopez, Garett",
    title = "{The CMB optical depth constrains the duration of reionization}",
    eprint = "2505.15899",
    archivePrefix = "arXiv",
    primaryClass = "astro-ph.CO",
    year = "2025"
}

@article{Sailer:2025xcy,
    author = "Sailer, Noah and Farren, Gerrit S. and Ferraro, Simone and White, Martin",
    title = "{Addressing Tensions in {\ensuremath{\Lambda}}CDM Cosmology by an Increase in the Optical Depth to Reionization}",
    eprint = "2504.16932",
    archivePrefix = "arXiv",
    primaryClass = "astro-ph.CO",
    doi = "10.1103/6r54-8lv4",
    journal = "Phys. Rev. Lett.",
    volume = "136",
    number = "8",
    pages = "081002",
    year = "2026"
}

@article{Jhaveri:2025neg,
    author = "Jhaveri, Tanisha and Karwal, Tanvi and Hu, Wayne",
    title = "{Turning a negative neutrino mass into a positive optical depth}",
    eprint = "2504.21813",
    archivePrefix = "arXiv",
    primaryClass = "astro-ph.CO",
    doi = "10.1103/6vd2-rbfn",
    journal = "Phys. Rev. D",
    volume = "112",
    number = "4",
    pages = "043541",
    year = "2025"
}

@article{Reichardt:2020jrr,
    author = "Reichardt, C. L. and others",
    title = "{An Improved Measurement of the Secondary Cosmic Microwave Background Anisotropies from the SPT-SZ + SPTpol Surveys}",
    eprint = "2002.06197",
    archivePrefix = "arXiv",
    primaryClass = "astro-ph.CO",
    doi = "10.3847/1538-4357/abfd98",
    journal = "Astrophys. J.",
    volume = "908",
    pages = "199",
    year = "2021"
}

@article{Louis:2025tst,
    author = "Louis, Thibaut and others",
    collaboration = "ACT",
    title = "{The Atacama Cosmology Telescope: DR6 CMB power spectra and likelihood}",
    eprint = "2503.14452",
    archivePrefix = "arXiv",
    primaryClass = "astro-ph.CO",
    year = "2025"
}

@article{MacCrann:2024ato,
    author = "MacCrann, Niall and others",
    collaboration = "ACT",
    title = "{The Atacama Cosmology Telescope: Reionization kSZ trispectrum methodology and limits}",
    eprint = "2405.01188",
    archivePrefix = "arXiv",
    primaryClass = "astro-ph.CO",
    doi = "10.1093/mnras/stae1607",
    journal = "Mon. Not. Roy. Astron. Soc.",
    volume = "532",
    pages = "4247",
    year = "2024"
}

@article{Chen:2022cgw,
    author = "Chen, Nianyi and Trac, Hy and Mukherjee, Suvodip and Cen, Renyue",
    title = "{Patchy kinetic Sunyaev--Zel'dovich effect with controlled reionization history and morphology}",
    eprint = "2203.04337",
    archivePrefix = "arXiv",
    primaryClass = "astro-ph.CO",
    doi = "10.3847/1538-4357/acb3a7",
    journal = "Astrophys. J.",
    volume = "943",
    pages = "138",
    year = "2023"
}

@article{Trac:2021ctk,
    author = "Trac, Hy and Chen, Nianyi and Holber, Ian and Cen, Renyue and Zuber, Justas",
    title = "{AMBER: A Semi-numerical Abundance Matching Box for the Epoch of Reionization}",
    eprint = "2109.10375",
    archivePrefix = "arXiv",
    primaryClass = "astro-ph.CO",
    doi = "10.3847/1538-4357/ac5116",
    journal = "Astrophys. J.",
    volume = "927",
    pages = "186",
    year = "2022"
}

@article{Battaglia:2012im,
    author = "Battaglia, Nicholas and Natarajan, Adi and Trac, Hy and Cen, Renyue and Loeb, Abraham",
    title = "{Reionization on Large Scales III: Predictions for Low-$\ell$ CMB Polarization from High Resolution Simulations of Patchy Reionization}",
    eprint = "1211.2832",
    archivePrefix = "arXiv",
    primaryClass = "astro-ph.CO",
    doi = "10.1088/0004-637X/776/2/83",
    journal = "Astrophys. J.",
    volume = "776",
    pages = "83",
    year = "2013"
}

@article{Shull:2011aa,
    author = "Shull, J. Michael and Harness, Aparna and Trenti, Michele and Smith, Britton D.",
    title = "{Critical Star Formation Rates for Reionization: Full Reionization Occurs at Redshift $z \approx 7$}",
    eprint = "1108.3334",
    archivePrefix = "arXiv",
    primaryClass = "astro-ph.CO",
    doi = "10.1088/0004-637X/747/2/100",
    journal = "Astrophys. J.",
    volume = "747",
    pages = "100",
    year = "2012"
}

@article{Visbal:2018vfh,
    author = "Visbal, Eli and Haiman, Zoltan and Bryan, Greg L.",
    title = "{Self-consistent semi-analytic models of the first stars}",
    eprint = "1705.09005",
    archivePrefix = "arXiv",
    primaryClass = "astro-ph.CO",
    doi = "10.1093/mnras/sty142",
    journal = "Mon. Not. Roy. Astron. Soc.",
    volume = "475",
    pages = "5246-5256",
    year = "2018"
}

@article{Slatyer:2016qyl,
    author = "Slatyer, Tracy R.",
    title = "{Indirect dark matter signatures in the cosmic dark ages. II. Ionization, heating, and photon spectrum from arbitrary energy injections}",
    eprint = "1506.03812",
    archivePrefix = "arXiv",
    primaryClass = "hep-ph",
    doi = "10.1103/PhysRevD.93.023521",
    journal = "Phys. Rev. D",
    volume = "93",
    pages = "023521",
    year = "2016"
}

@article{Foreman-Mackey:2012any,
    author = "Foreman-Mackey, Daniel and Hogg, David W. and Lang, Dustin and Goodman, Jonathan",
    title = "{emcee: The MCMC Hammer}",
    eprint = "1202.3665",
    archivePrefix = "arXiv",
    primaryClass = "astro-ph.IM",
    doi = "10.1086/670067",
    journal = "Publ. Astron. Soc. Pac.",
    volume = "125",
    pages = "306--312",
    year = "2013"
}

@article{Zhu:2023dqs,
    author = "Zhu, Yongda and others",
    title = "{Probing Ultralate Reionization: Direct Measurements of the Mean Free Path over 5 {\ensuremath{<}} z {\ensuremath{<}} 6}",
    eprint = "2308.04614",
    archivePrefix = "arXiv",
    primaryClass = "astro-ph.CO",
    doi = "10.3847/1538-4357/aceef4",
    journal = "Astrophys. J.",
    volume = "955",
    number = "2",
    pages = "115",
    year = "2023"
}

@article{Becker:2024ybt,
    author = "Becker, George D. and Bolton, James S. and Zhu, Yongda and Hashemi, Seyedazim",
    title = "{Damping wing absorption associated with a giant Ly{\,}{\ensuremath{\alpha}} trough at z {\ensuremath{<}} 6: direct evidence for late-ending reionization}",
    eprint = "2405.08885",
    archivePrefix = "arXiv",
    primaryClass = "astro-ph.CO",
    doi = "10.1093/mnras/stae1918",
    journal = "Mon. Not. Roy. Astron. Soc.",
    volume = "533",
    number = "2",
    pages = "1525--1540",
    year = "2024"
}

@article{PlanckIntXLVII,
    author = "Adam, R. and others",
    collaboration = "Planck",
    title = "{Planck Intermediate Results. XLVII. Constraints on reionization history}",
    eprint = "1605.03507",
    archivePrefix = "arXiv",
    primaryClass = "astro-ph.CO",
    doi = "10.1051/0004-6361/201628897",
    journal = "Astron. Astrophys.",
    volume = "596",
    pages = "A108",
    year = "2016"
}

@article{Smith:2026skb,
    author = "Smith, Aaron and others",
    title = "{The Lumina Project: CMB Optical Depth Fluctuations from Patchy Reionization}",
    eprint = "2605.18939",
    archivePrefix = "arXiv",
    primaryClass = "astro-ph.CO",
    month = "5",
    year = "2026"
}

@article{Hutter:2026kxp,
    author = "Pietschke, Yannic and Hutter, Anne and Heneka, Caroline",
    title = "{Constraining Reionization Morphology and Source Properties with 21cm-Galaxy Cross-Correlation Surveys}",
    eprint = "2601.18627",
    archivePrefix = "arXiv",
    primaryClass = "astro-ph.CO",
    month = "1",
    year = "2026"
}

@article{Kageura:2026opt,
    author = "Kageura, Akira and others",
    title = "{A New Constraint on the Optical Depth from the Reionization History Independent of CMB Large-Scale E-Mode Polarization}",
    eprint = "2601.09644",
    archivePrefix = "arXiv",
    primaryClass = "astro-ph.CO",
    year = "2026"
}

@article{Torrado:2020dgo,
    author = "Torrado, Jes\'us and Lewis, Antony",
    title = "{Cobaya: Code for Bayesian Analysis of hierarchical physical models}",
    eprint = "2005.05290",
    archivePrefix = "arXiv",
    primaryClass = "astro-ph.IM",
    doi = "10.1088/1475-7516/2021/05/057",
    journal = "JCAP",
    volume = "05",
    pages = "057",
    year = "2021"
}

@article{Blas:2011rf,
    author = "Blas, Diego and Lesgourgues, Julien and Tram, Thomas",
    title = "{The Cosmic Linear Anisotropy Solving System (CLASS) II: Approximation schemes}",
    eprint = "1104.2933",
    archivePrefix = "arXiv",
    primaryClass = "astro-ph.CO",
    doi = "10.1088/1475-7516/2011/07/034",
    journal = "JCAP",
    volume = "07",
    pages = "034",
    year = "2011"
}

@article{Sethi:2004pe,
    author = "Sethi, S. K. and Subramanian, K.",
    title = "{Primordial magnetic fields in the post-recombination era and early reionization}",
    journal = "Mon. Not. Roy. Astron. Soc.",
    volume = "356",
    pages = "778",
    year = "2005",
    eprint = "astro-ph/0405413",
    archivePrefix = "arXiv",
    doi = "10.1111/j.1365-2966.2004.08520.x"
}

@article{Kunze:2013uja,
    author = "Kunze, Kerstin E. and Komatsu, Eiichiro",
    title = "{Constraining primordial magnetic fields with distortions of the black-body spectrum of the cosmic microwave background: pre- and post-decoupling contributions}",
    journal = "JCAP",
    volume = "01",
    pages = "009",
    year = "2014",
    eprint = "1309.7994",
    archivePrefix = "arXiv",
    primaryClass = "astro-ph.CO",
    doi = "10.1088/1475-7516/2014/01/009"
}

@article{Elbers:2025mho,
    author = "Elbers, Willem and others",
    collaboration = "DESI",
    title = "{Constraints on Neutrino Physics from DESI DR2 BAO and DR1 Full Shape}",
    eprint = "2503.14744",
    archivePrefix = "arXiv",
    primaryClass = "astro-ph.CO",
    journal = "Phys. Rev. D",
    volume = "112",
    pages = "083513",
    year = "2025"
}

@article{Hazumi:2019lub,
    author = "Hazumi, M. and others",
    collaboration = "LiteBIRD",
    title = "{LiteBIRD: JAXA's new strategic L-class mission for all-sky surveys of cosmic microwave background polarization}",
    eprint = "2101.12449",
    archivePrefix = "arXiv",
    primaryClass = "astro-ph.IM",
    doi = "10.1117/12.2563050",
    journal = "Proc. SPIE Int. Soc. Opt. Eng.",
    volume = "11443",
    pages = "114432F",
    year = "2020"
}

@article{Ali-Haimoud:2016mbv,
    author = "Ali-Haimoud, Yacine and Kamionkowski, Marc",
    title = "{Cosmic microwave background limits on accreting primordial black holes}",
    eprint = "1612.05644",
    archivePrefix = "arXiv",
    primaryClass = "astro-ph.CO",
    doi = "10.1103/PhysRevD.95.043534",
    journal = "Phys. Rev. D",
    volume = "95",
    number = "4",
    pages = "043534",
    year = "2017"
}

@article{Abazajian:2019eic,
    author = "Abazajian, Kevork and others",
    title = "{CMB-S4 Science Case, Reference Design, and Project Plan}",
    eprint = "1907.04473",
    archivePrefix = "arXiv",
    primaryClass = "astro-ph.IM",
    reportNumber = "FERMILAB-PUB-19-431-AE-SCD",
    month = "7",
    year = "2019"
}

@article{Robertson:2015uda,
    author = "Robertson, Brant E. and Ellis, Richard S. and Furlanetto, Steven R. and Dunlop, James S.",
    title = "{Cosmic Reionization and Early Star-forming Galaxies: a Joint Analysis of new Constraints From Planck and the Hubble Space Telescope}",
    eprint = "1502.02024",
    archivePrefix = "arXiv",
    primaryClass = "astro-ph.CO",
    doi = "10.1088/2041-8205/802/2/L19",
    journal = "Astrophys. J. Lett.",
    volume = "802",
    number = "2",
    pages = "L19",
    year = "2015"
}

@article{DESI:2024mwx,
    author = "Adame, A. G. and others",
    collaboration = "DESI",
    title = "{DESI 2024 VI: cosmological constraints from the measurements of baryon acoustic oscillations}",
    eprint = "2404.03002",
    archivePrefix = "arXiv",
    primaryClass = "astro-ph.CO",
    reportNumber = "FERMILAB-PUB-24-0154-PPD",
    doi = "10.1088/1475-7516/2025/02/021",
    journal = "JCAP",
    volume = "02",
    pages = "021",
    year = "2025"
}

@ARTICLE{2011arXiv1104.2934L,
       author = {{Lesgourgues}, Julien},
        title = "{The Cosmic Linear Anisotropy Solving System (CLASS) III: Comparision with CAMB for LambdaCDM}",
      journal = {arXiv e-prints},
         year = 2011,
        month = apr,
          eid = {arXiv:1104.2934},
        pages = {arXiv:1104.2934},
          doi = {10.48550/arXiv.1104.2934},
archivePrefix = {arXiv},
       eprint = {1104.2934},
 primaryClass = {astro-ph.CO},
       adsurl = {https://ui.adsabs.harvard.edu/abs/2011arXiv1104.2934L}
}

@article{Das:2019sda,
    author = "Lewis, Antony and Challinor, Anthony and Lasenby, Anthony",
    title = "{Efficient computation of CMB anisotropies in closed FRW models}",
    eprint = "astro-ph/9911177",
    archivePrefix = "arXiv",
    primaryClass = "astro-ph.CO",
    reportNumber = "FERMILAB-PUB-19-572-A",
    doi = "10.1086/309179",
    journal = "Astrophys. J.",
    volume = "538",
    pages = "473",
    year = "2000"
}

@article{Speagle:2020,
  author = "Speagle, Joshua S.",
    title  = "{dynesty: a dynamic nested sampling package for estimating Bayesian posteriors and evidences}",
      journal = "Mon. Not. Roy. Astron. Soc.",
        volume = "493", pages = "3132", year = "2020",
          eprint = "1904.02180", archivePrefix = "arXiv", primaryClass = "astro-ph.IM"
          }

\end{document}